\documentclass[a4paper,12pt]{article}
\usepackage[papersize={216mm,330mm},tmargin=15mm,bmargin=15mm,lmargin=15mm,rmargin=15mm]{geometry}
\usepackage{fancyhdr}
\usepackage{amsmath}
\usepackage{amssymb}
\usepackage{xcolor}
\usepackage{graphicx,multicol}
\usepackage{subfigure}
\usepackage{bm}
\usepackage{tabu}
\usepackage{hyperref}
\usepackage{textgreek}
\usepackage{upgreek}
\usepackage{multicol}
\usepackage{natbib}
\usepackage{comment}
\usepackage{setspace}
\usepackage{siunitx}
\usepackage[utf8]{inputenc}
\usepackage[T1]{fontenc}
\usepackage{lineno}
\usepackage{appendix}
\usepackage{mathrsfs}
\setcitestyle{authoryear}

\title{Analytic model for feature maps in the primary visual cortex}

\usepackage{authblk}
\author[1,2]{X. Liu}
\author[1,2]{P.A. Robinson}

\affil[1]{School of Physics, University of Sydney, New South Wales 2006, Australia}
\affil[2]{Center for Integrative Brain Function, University of Sydney, New South Wales 2006, Australia}

\begin{document}
\maketitle

\begin{abstract}
A compact analytic model is proposed to describe the combined orientation preference (OP) and ocular dominance (OD) features of simple cells and their layout in the primary visual cortex (V1). This model consists of three parts: (i) an anisotropic Laplacian (AL) operator that represents the local neural sensitivity to the orientation of visual inputs; (ii) a receptive field (RF) operator that models the anisotropic spatial RF that projects to a given V1 cell over scales of a few tenths of a millimeter and combines with the AL operator to give an overall OP operator; and (iii) a map that describes how the parameters of these operators vary approximately periodically across V1. The parameters of the proposed model maximize the neural response at a given OP with an OP tuning curve fitted to experimental results. It is found that the anisotropy of the AL operator does not significantly affect OP selectivity, which is dominated by the RF anisotropy, consistent with Hubel and Wiesel's original conclusions that orientation tuning width of V1 simple cell is inversely related to the elongation of its RF.  A simplified OP-OD map is then constructed to describe the approximately periodic OP-OD structure of V1 in a compact form. Specifically, the map is approximated by retaining its dominant spatial Fourier coefficients, which are shown to suffice to reconstruct the overall structure of the OP-OD map. This representation is a suitable form to analyze observed maps compactly and to be used in neural field theory of V1. Application to independently simulated V1 structures shows that observed irregularities in the map correspond to a spread of dominant coefficients in a circle in Fourier space.
\end{abstract}

\maketitle
\doublespacing
\thispagestyle{empty}

\section{Introduction}
\label{sec:Intro}
V1 is the first cortical area that processes visual inputs from the lateral geniculate nucleus (LGN) of the thalamus before projecting output signals to higher visual areas \citep{Miikkulainen_visual_maps}. The feedforward visual pathway from the eyes to V1 involves two main processing steps: (i) light levels at a given spatial location are detected and converted into neural signals by the retina ganglion cells; and (ii) the neural signals are transmitted to V1 through the lateral geniculate nuclei (LGN) of the thalamus \citep{schiller_visual}. LGN neurons have approximately circular receptive fields with either a central ON region (activity enhanced by light incident there) surrounded by an OFF annulus (activity suppressed by light incident there), or vice versa \citep{deangelis1995receptive,hubel1961integrative}.

Similarly to other parts of the cortex, V1 can be approximated as a two-dimensional sheet when studying the spatial structure of the retinotopic map \citep{Tovee_visual_system}. V1 neurons, which respond to same eye preference or orientation preference, are arranged in columns perpendicular to the cortical surface. Columns do not have sharp boundaries; rather, feature preferences gradually vary across the surface of V1. These maps are overlaid such that a given neuron responds to several features \citep{Hubel_column_arrangm_1962, hubel1968receptive,hubel_sequence_1974,Miikkulainen_visual_maps}. 

\begin{figure}[h!]
\centering
\includegraphics[width=0.55\textwidth]{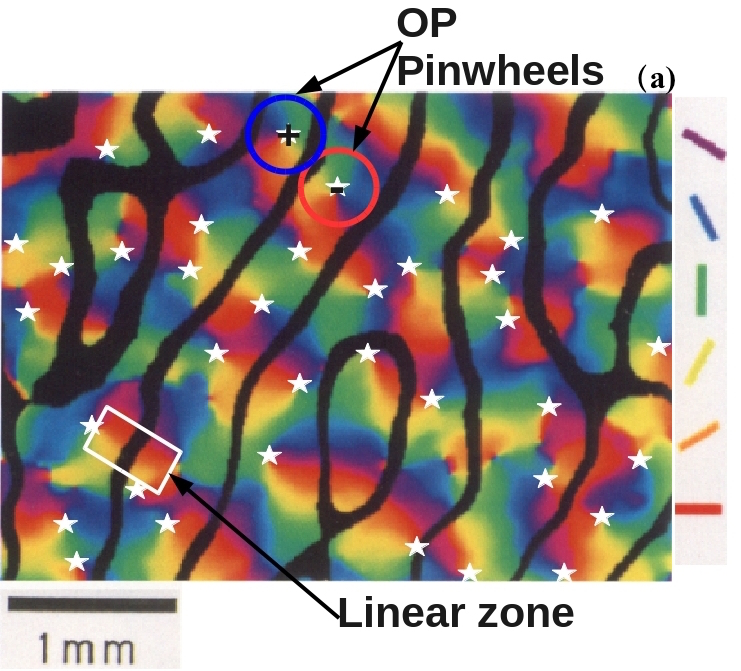}
\includegraphics[width=0.5 \textwidth]{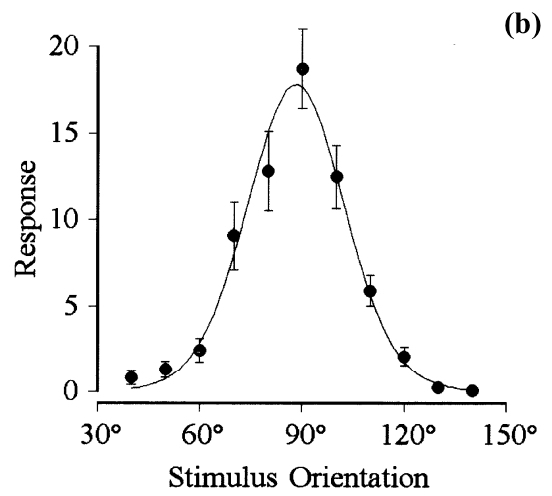}
\caption{(a) Combined OP-OD map of macaque monkey, adapted from \cite{blasdel_orientation_1992}. The borders of OD stripes are shown in solid black, and singularities (pinwheel centers) are labeled by white stars. Oriented color bars indicate different OPs. The blue and red circles outline examples of positive and negative OP pinwheels, and the white rectangle outlines a linear zone. (b) Experimental orientation tuning curve, adapted from \cite{Swindale1998}. The preferred orientation angel is around $90^\circ$. The dots are the data points, and the solid curve is the fitted tuning curve using a von Mises function.}
\label{fig:op_od_mix}
\end{figure}

Two prominent feature preferences of V1 cells are their layout in a combined the OP-OD map, as seen in Fig.~\ref{fig:op_od_mix}(a), which shows an example from experiment \citep{blasdel_orientation_1992}. \cite{hubel1968receptive} found that neurons that  respond preferentially to stimuli from one eye or the other are arranged in alternating bands across layer 4C of V1 in macaque monkeys, and these bands are termed left and right OD columns. The average OD column width in mammals ranges from $\sim 0.5 - 1$ mm  \citep{adams_complete_2007,Horton_OD}. An OP column, sometimes called an iso-orientation slab, comprises neurons that respond to similar edge orientation in a visual field. Each OP column not only spans several cortical layers vertically, but also extends $25 - 50\  \mu$m laterally in monkey. Moreover, OP normally varies continuously as a function of the cortical position, covering the complete range  $0^\circ$ to $180^\circ$ of edge orientations  \citep{hubel_sequence_1974,hubel_ferrier_1977,obermayer_geometry_1993}. Optical imaging  reveals that OP columns are quasiperiodic, and are arranged as pinwheels, within which each of the OPs varies azimuthally around a center called a singularity \citep{blasdel_orientation_1992,bonhoeffer_iso-orientation_1991,bonhoeffer_layout_1993, swindale_review_1996}. Furthermore, the OP in each pinwheel increases either clockwise (negative pinwheel) or counterclockwise (positive pinwheel) and most neighboring pinwheels have opposite signs \citep{gotz_d-blob_1987,Gotz_1988}. 
Examples of positive and negative OP pinwheels are outlined in Fig.~\ref{fig:op_od_mix}(a). 
The superimposed OD and OP maps have specific relationships, including that: (i) most pinwheels are centered near the middle of OD stripes; (ii) linear zones, which are formed by near-parallel OP columns, usually connect two singularities and cross the border of OD stripes at right angles \citep{bartfeld_relationships_1992}, as highlighted in the white rectangle in Fig.~\ref{fig:op_od_mix}(a). Additionally, various studies \citep{chklovskii2004maps,koulakov2001orientation,mitchison1991neuronal,mitchison1995type} have argued that the appearance of the OP-OD map reflects wiring optimization of local neuron connectivity, in which the distance between neurons with similar feature preference is kept as small as possible. 

According to the quasiperiodicity of the feature preference of V1, previous studies \citep{bressloff_visual_2002,Veltz_2015} suggested that the functional
maps of V1 can be approximated by a spatially periodic network of fundamental domains, each of which is called a hypercolumn. Each hypercolumn represents a small piece of V1, which consists of left and right OD stripes with a pair of positive and negative pinwheels in each, so as to ensure the complete coverage of the OP and OD selectivity. 

Orientation selectivity plays a primary role in early-stage visual processing. One way to characterize the preferred orientations of a single neuron is to measure the tuning curve from its neuronal response to visual stimuli with various orientations. Figure~\ref{fig:op_od_mix}(b) shows an experimental orientation tuning curve obtained from single unit responses in area 17 of adult cat, with optimal orientation angle of $90^\circ$ \citep{Swindale1998}. A typical full width at half maximum (FWHM) of such a curve is $\sim35^\circ$.

\begin{figure}[h!]
\centering
\includegraphics[width=0.75 \textwidth]{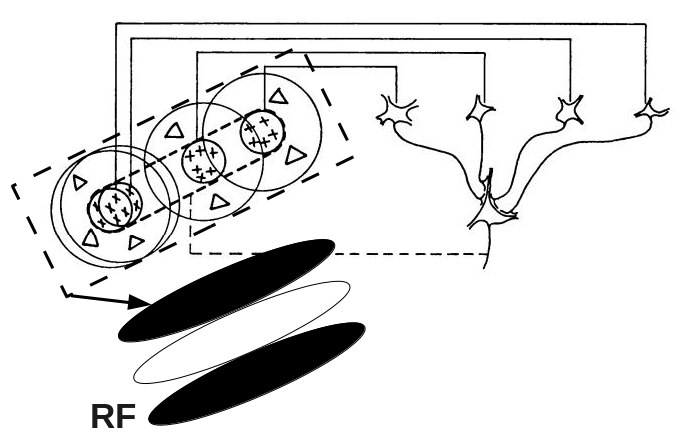}
\caption{Schematic of the elongated RF of a V1 simple cell, showing  the convergence of several LGN RFs into the V1 RF.  The four cells in the right half of the figure represents LGN cells with circular ON center, OFF surround RFs. The outputs of these LGN cells project to form the elongated V1 RF shown at the bottom left. Adapted from \cite{hubel1962receptive}. }
 \label{fig:V1_LGN_RF}
\end{figure}

 The mapping of inputs from the retina to V1 is organized in a retinotopic manner. Visual information in nearby regions within the visual field is projected to neighboring ganglion cells in the retina. This spatial arrangement is maintained through the LGN to V1, where the visual signals are further processed by neighboring V1 neurons. The subregion in the visual field, within which certain features of the visual object tend to evoke or suppress neural firing of a given V1 neuron, is termed the receptive field (RF) of that neuron \citep{hubel1962receptive,schiller_visual,skottun1991classifying,smith_RF_size,roger_retinotopic}. This paper focuses on the RF of V1 simple cells, which respond best to oriented bars. The spatial RF of a V1 simple cell has separate ON (excitatory) and OFF (inhibitory) subareas, which are elongated in a specific orientation, and these subareas relate to the ON and OFF regions of LGN RFs that project to the V1 RF of a given cell. The neuron will be excited when light illuminates the ON subarea, and be depressed when light exposed to the OFF subarea  \citep{hubel1968receptive,MECHLER20021017}. It was first proposed by \cite{hubel1962receptive} that the RF of a V1 simple cell can be predicted by a feed-forward model. Specifically, they suggested that RF of a V1 simple cell is formed by by combining the circular RFs of several LGN cells to produce an elongated RF with central ON region flanked by OFF regions. Figure \ref{fig:V1_LGN_RF} shows a schematic of this feed-forward model, which produces a RF with a three-lobed pattern, as shown in the bottom left corner.  In addition, several studies have suggested that the lateral intracortical excitatory and inhibitory connectivities from surrounding neurons also play important roles in a cell's orientation tuning and RF formation \citep{ferster2000neural, finn2007emergence,gardner1999linear, marino2005invariant,Moore2012DevelopmentOO}. Moreover, some studies have discussed the relationship between the size of the RF and the width of the orientation tuning curve \citep{hubel1962receptive, lampl2001prediction}. These authors predicted that the width of orientation tuning curve should be inversely associated with the elongation of the RF.

Numerous experiments have used optical imaging or functional magnetic resonance imaging (fMRI) to reveal the spatial structure of the OP-OD map in mammals including humans \citep{bartfeld_relationships_1992,blasdel_orientation_1992,bonhoeffer_iso-orientation_1991,bonhoeffer_layout_1993,bosking_orientation_1997,obermayer_geometry_1993,yacoub_human_op_map}. Additionally, a number of neural network models have been proposed for the development of OP-OD maps and simulated numerically \citep{bednar2009topographica,erwin_models_1995,obermayer1992model,swindale_model_1992,Miikkulainen_visual_maps,Stevens_GCAL}. The resulting OP-OD maps obtained from experiments or simulation are only semiregular, as illustrated in Fig.~\ref{fig:op_od_mix}(a). Hence, it usually requires many data points to describe the structure of such maps, which impedes understanding and requires extensive computation to integrate OP-OD maps into models of neural activity in V1. Such models would benefit from a compact analytic representation of the OP-OD map; e.g., to incorporate its structure into existing spatiotemporal correlation analyses of gamma-band oscillations of neural activity using neural field theory (NFT), or to understand propagation via patchy neural connections in V1  \citep{robinson_propagator_2005,robinson_patchy_2006,robinson_visual_2007, Lxc2020gamma_correlation}. These issues motivate us to derive a compact analytical representation of the OP-OD map for use in prediction and analysis of brain activity, and for  analysis of properties of OP maps obtained from \textit{in-vivo} experiments or computer simulations.

To achieve the above aim, we first note that it has long been suggested that oriented visual features such as edges can be detected by the Laplacian operator \citep{marr1982vision, marr1980theory,marr1981directional,machband1965,young1987gaussian}. Hence, we approximate the local sensitivity of V1 neuron to the orientation of stimuli by such an operator, which we allow to be anisotropic. This operator incorporates the details of LGN receptive fields, as projected to the cortex, and any anisotropic response at V1, as implemented through local excitatory and inhibitory wiring between neurons. Secondly, we introduce an RF operator to approximate the anisotropy of the visual region that projects activity to a given V1 simple cell. The combined Laplacian and RF operators yield an overall OP operator at each point, whose parameters can be fitted to yield observed OP tuning widths. Thirdly, we allow the properties of the OP operator to vary across V1 approximately periodically, as described by dominant Fourier coefficients.

The paper is structured as follows: Section~\ref{sec:unitcell} approximates the hypercolumn with its structure being compatible with the general features of OP-OD map. Meanwhile, it also is compatible with NFT. In Sec.~\ref{sec:decomp_op}, we describe the local sensitivity to stimulus orientation by applying an anisotropic Laplacian operator to the stimulus. A RF operator is then introduced to project activity from neighboring neurons to a given point. The resulting combined OP operator maximizes the response of V1 neurons to their preferred stimulus orientation and its parameters are adjusted to match experimental tuning curves. In Sec.~\ref{sec:fourier_analysis}, we Fourier decompose the OP variation on the period of an idealized hypercolumn and investigate the properties of the resulting Fourier coefficients. The results are then applied to compactly represent and analyze OP maps generated from widely used simulation models are then investigated in Sec.~\ref{sec:applications}. Finally, the results are discussed and summarized in Sec.~\ref{sec:summary}.

\section{Hypercolumns and OD-OP maps} 
\label{sec:unitcell}
In this section, an approximate OP-OD map within a hypercolumn is proposed, which reproduces the main aspects of observed maps. It is Fourier decomposed in later sections to generate a sparse set of Fourier coefficients that can be used in NFT to study activity across many hypercolumns.

\subsection{Hypercolumn arrangement}
\label{sec: unitcell_arrangemt}
\begin{figure}[h!]
\centering
\includegraphics[width=.4\textwidth]{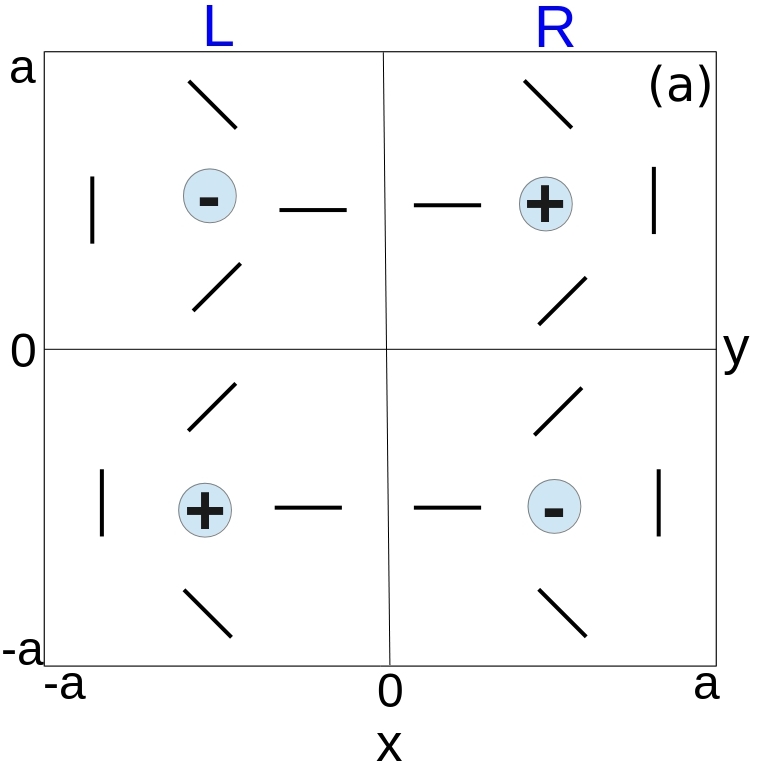}%
\includegraphics[width=.4\textwidth]{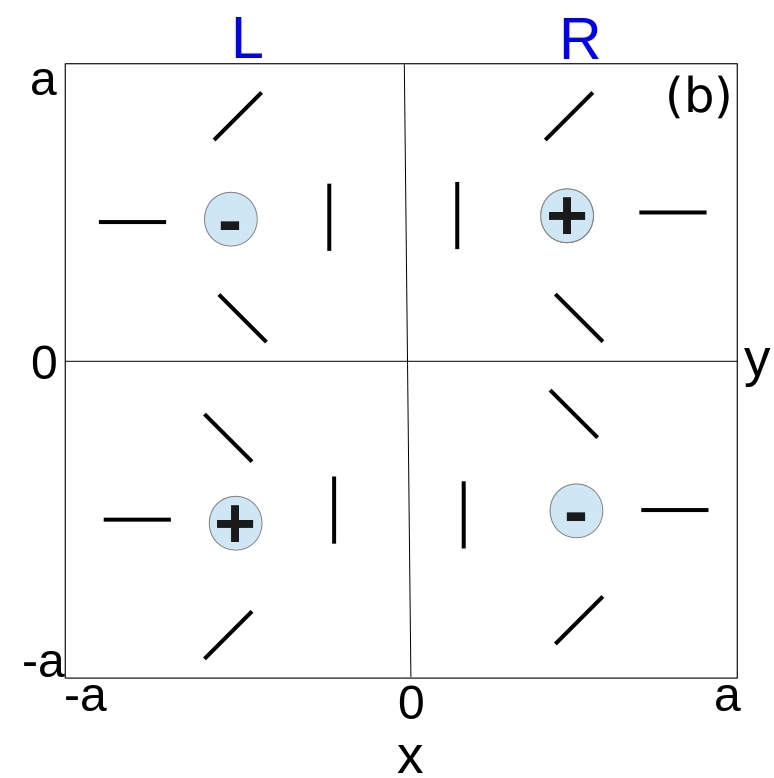}
\caption{Schematics of possible hypercolumn arrangements. The two vertical columns of each hypercolumn represent the left and right OD bands. The orientated bars represent the OP within a pinwheel, and the $+/-$ signs indicate the polarity of the pinwheels. (a) Pinwheel arrangement I. (b) Pinwheel arrangement II.}
\label{fig:pin_arrgmt}
\end{figure}

All feature preferences within a small visual field are mapped to a hypercolumn in V1 \citep{Hubel_column_arrangm_1962,hubel_sequence_1974,Miikkulainen_visual_maps}. Based on the pinwheel model introduced by \cite{devalois1990spatial}, we approximate the hypercolumn as a square domain, which consists of left and right OD stripes of equal width; each OD stripe contains a pair of positive and negative pinwheels for continuous (except at pinwheel centers) feature preference coverage within a hypercolumn.
The hypercolumn is consistent with general observations of visual cortical map formation from experimental studies \citep{adams_complete_2007,bartfeld_relationships_1992,bonhoeffer_iso-orientation_1991, erwin_models_1995,muller_analysis_2000,obermayer_statistical-mechanical_1992,obermayer_geometry_1993}, including that: (i) left-eye and right-eye OD columns are arranged as alternating stripes in V1 of average width $a\approx 1$ mm; (ii) OP angles are arranged as pinwheels; (iii) each pinwheel center coincides with the center of its OD band; (iv) OP is continuous at OD boundaries; and (iv) neighboring pinwheels have opposite signs. A $+$ sign denotes a counterclockwise increase of OP around the pinwheel center, whereas a $-$ sign denotes a clockwise increase. According to the rules described above, two distinct arrangements of the hypercolumn are possible, as illustrated in Fig.~\ref{fig:pin_arrgmt}, where each hypercolumn is $2a$ wide. Pinwheel arrangement I in Fig.~\ref{fig:pin_arrgmt}(a) is used for further study in later sections, without loss of generality. However, other arrangements produce analogous results due to the fact that the hypercolumn are assumed to be continuous at boundary and periodic across V1, and other hypercolumn arrangements can be obtained by either rotating the pinwheels clockwise/counterclockwise or swapping the left/right column or top/bottom row horizontally or vertically of Fig.~\ref{fig:pin_arrgmt}(a). Note also that one is free to consider a hypercolumn whose lower left corner is at the center of the current hypercolumn in Fig.~\ref{fig:pin_arrgmt}(a). Such a hypercolumn has exactly the same structure as in Fig.~\ref{fig:pin_arrgmt}(b), except for interchange of the left- and right-eye columns.

\subsection{Single OP pinwheel structure} 
\label{pinwheel_visualize}
Our model determines the OP as a function of cortical location within each hypercolumn. The spatial coordinates of the hypercolumn are set by placing the origin of the coordinates at the center of the hypercolumn, whose boundaries are at $x=\pm a$ and $y=\pm a$, as shown in Fig.~\ref{fig:pin_arrgmt}. The four pinwheels in a single hypercolumn are modeled by first generating the right-top pinwheel, and other pinwheels are produced by mirroring the right-top pinwheel across the $x$-axis, the $y$-axis, and then both.
When generating the right-top pinwheel, the $x$ and $y$ coordinates range from $0$ to $a$ and the OP angle $\varphi(x,y)$ at each cortical position $(x,y)$ is approximated by the inverse tangent function defined in Eq.~\eqref{eq:inverse_tan_op_angle}.  
\begin{equation}
\label{eq:inverse_tan_op_angle}
\varphi(x,y)= \frac{1}{2}
\begin{cases}
    \arctan\left(\frac{y-y_{0}}{x-x_{0}}\right), & x>x_0, y>y_0 \\
    \frac{\pi}{2}, & x=x_0, y>y_0 \\
    \arctan\left(\frac{y-y_{0}}{x-x_{0}}\right)+ \pi, & x<x_0, y>y_0 \\
    \frac{3\pi}{2}, & x=x_0, y<y_0 \\
    \arctan\left(\frac{y-y_{0}}{x-x_{0}}\right)+ 2\pi, & x>x_0, y<y_0 
  \end{cases}
\end{equation}    
where $(x_{0},y_{0}) = (a/2, a/2)$ is the center of the right-top pinwheel. The $1/2$ coefficient in front of the inverse tangent functions is to make the range of $\varphi(x,y)$ to be $0^\circ$ to $180^\circ$.

\begin{figure}[h!]
\centering
\includegraphics[width=1.01 \textwidth]{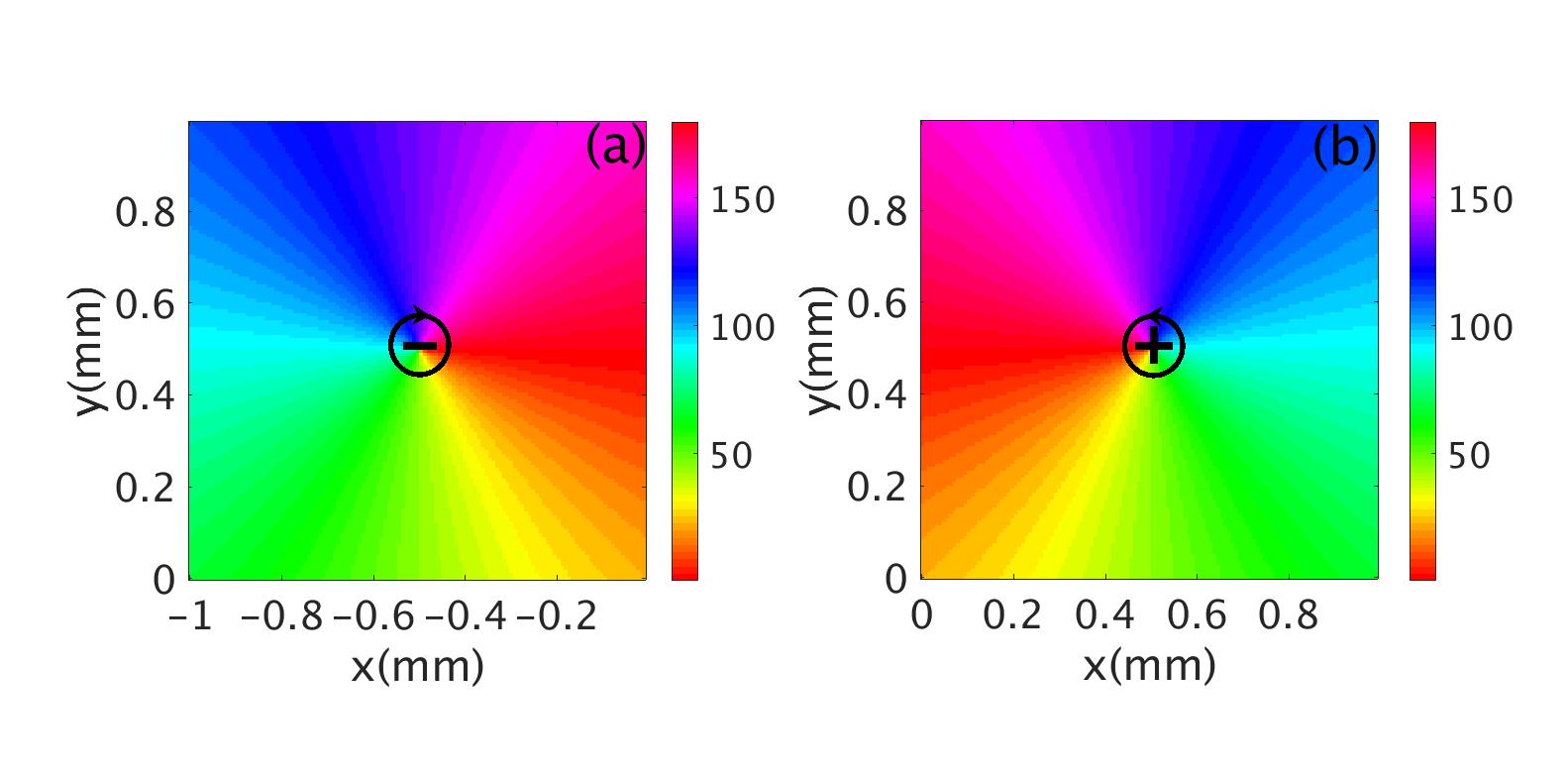}
\includegraphics[width=0.45\textwidth]{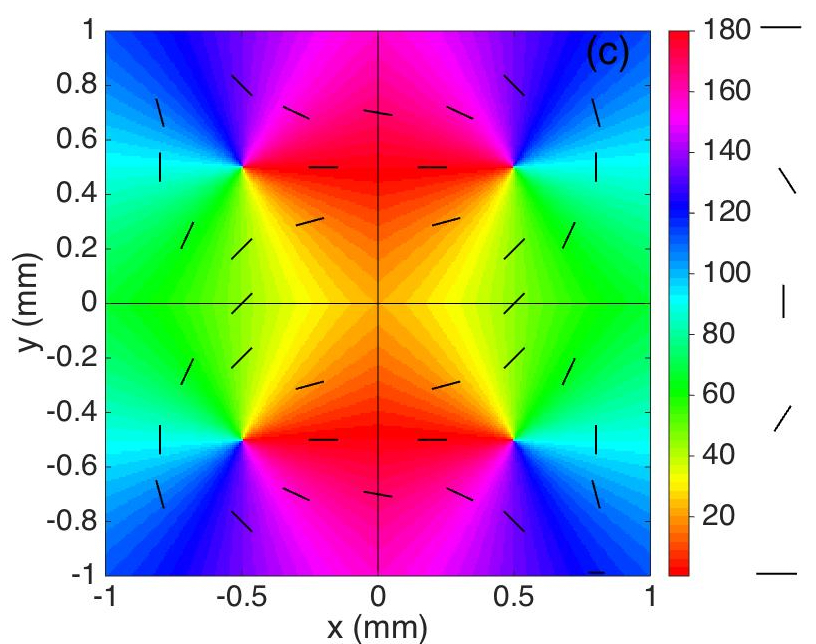}%
\includegraphics[width=0.4\textwidth]{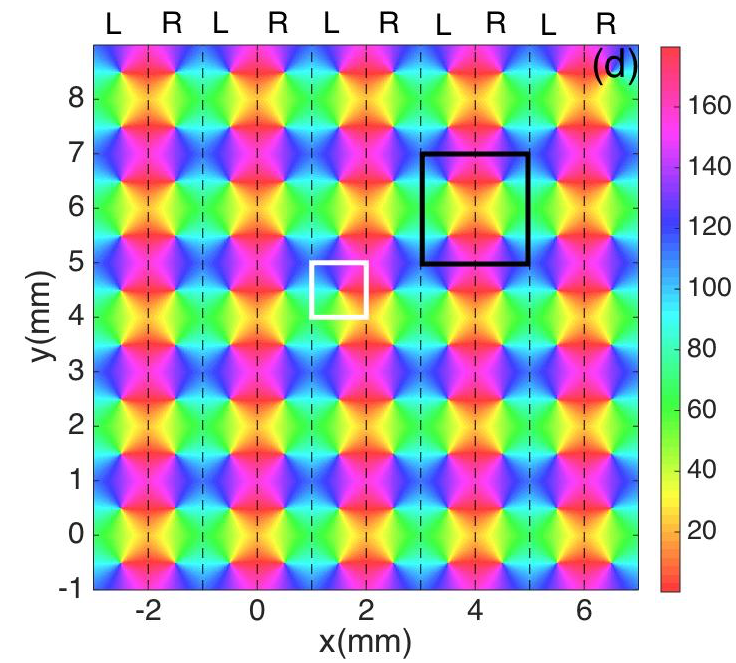}
\caption{Schematics of visual feature preference maps in V1 with color bars indicating OP in degrees. (a) Negative pinwheel. (b) Positive pinwheel. Both pinwheels are $180^\circ$ periodic. (c) Hypercolumn. The vertical line divides the hypercolumn into left and right OD columns of equal width, while the horizontal and vertical lines split the hypercolumn into four squares, each containing one OP pinwheel. The short bars highlight the OP at various locations. The $0^\circ$ and $180^\circ$ orientations both appear as horizontal bars. (d) Periodic spatial structure of OP and OD columns across a small piece of V1 comprising 25 hypercolumns. Dashed lines bound left (L) and right (R) OD columns. One pinwheel is outlined in white and one hypercolumn is outlined in black.}
 \label{fig:op_map}
\end{figure}

The negative and positive pinwheels on the top half of hypercolumn are illustrated in Figs~\ref{fig:op_map}(a) and (b). Figure~\ref{fig:op_map}(c) shows a hypercolumn containing four pinwheels. As mentioned previously, the OP and OD features are approximated as continuous and periodic for the moment, so V1 can be approximated as lattice of hypercolumns. Thus, we can construct an array of our approximated hypercolumns to represent a piece of V1, which is shown in Fig.~\ref{fig:op_map}(d). In such an array, the OP structure resembles maps reconstructed from \textit{in-vivo} experiments (e.g., Fig.~\ref{fig:op_od_mix}(a)), although the OD stripes are approximated as straight here \citep{blasdel_orientation_1992,bonhoeffer_iso-orientation_1991,bonhoeffer_layout_1993,obermayer_geometry_1993}.

\section{OP Operator}
\label{sec:decomp_op}
In this section, we derive analytic representations for the OP of V1 neurons. Firstly, we adopt an anisotropic Laplacian operator to describe the local response of the system to an edge, which can include both the near-isotropic response of the LGN plus any anisotropy introduced by local wiring in V1. We also introduce the RF operator that describes the projection of activity from nearby neurons in an anistropic surrounding region. The combination of these two operators gives an over OP operator, whose parameters we fit to match experimental tuning curves.

\subsection{Anisotropic Laplacian (AL) Operator}
\label{sec:op_operator}
\begin{figure}[h!]
\centering
\includegraphics[width=0.65 \textwidth]{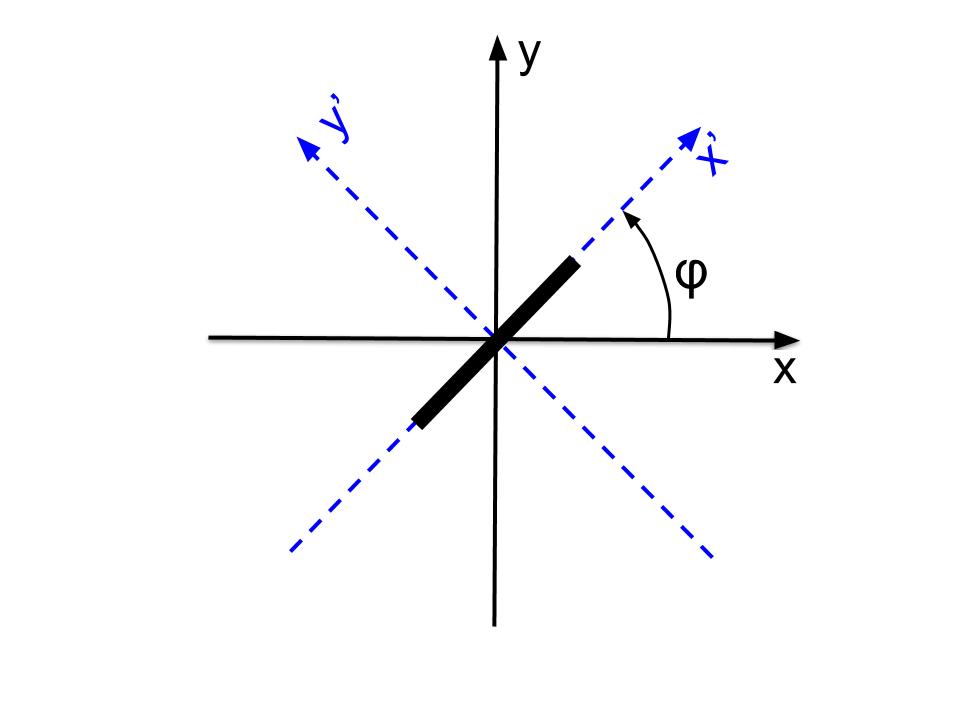}
\caption{Coordinates used to analyze the anisotropic Laplacian  operator. Original axes $x$ and $y$ are shown in solid, while the rotated axes $x'$ and $y'$ are dashed. The input stimulus is a bar oriented at angle $\varphi$.}
\label{fig:schematic_of_axes_rotation}
\end{figure}

In computer vision, edges with different orientation can be detected by linearly combining the second-order partial derivatives of the input \citep{marr1980theory, marr1982vision, torre1986edge}. Similarly, we can model the local sensitivity of V1 neurons to stimulus orientation, by calculating the weighted sum of the second-order partial derivatives of activity projected from the oriented bar stimuli, using rotated axes for simplicity. Figure~\ref{fig:schematic_of_axes_rotation} shows an original $x-y$ coordinate system, and $x'-y'$ axes obtained by rotating the original axes by an OP angle $\varphi$ that is the angle of an oriented bar.

A short bar in an image gives rise to localized, anisotropic intensity changes \citep{torre1986edge}. In the rotated coordinates, this 2-dimensional intensity change can be detected by the weighted second order partial derivatives in the $x'$ and $y'$ directions. Hence, we define an anisotropic Laplacian operator $\mathscr{P}$ as a weighted linear combination of the second order partial derivatives:
\begin{equation}
\label{eq:second_order_derivative_operator}
\mathscr{P}=a^2\frac{\partial^2 }{\partial x'^2}+b^2\frac{\partial^2 }{\partial y'^2}\,,
\end{equation}
where the rotated coordinates satisfy
\begin{equation}
x'=x \cos{\varphi}+y \sin{\varphi}\,,
\end{equation}
\begin{equation}
y'=-x \sin{\varphi}+ y \cos{\varphi}\,,
\end{equation}
where the OP $\varphi$ ranges from $0$ to $\pi$, and $a^2$ and $b^2$ are constants.

Taking the Fourier transform of both sides of Eq.~\eqref{eq:second_order_derivative_operator} yields
\begin{equation}
\label{eq:second_order_derivative_operator_FT}
\mathscr{P}(\mathbf{k})=-a^2k_{x'}^2-b^2k_{y'}^2\,,
\end{equation}
where
\begin{equation}
\label{k_x_prime}
k_{x'}^2=(k_{x}\cos{\varphi}+k_{y}\sin{\varphi})^2\,,
\end{equation}
\begin{equation}
\label{k_y_prime}
k_{y'}^2=(-k_{x}\sin{\varphi}+ k_{y}\cos{\varphi})^2\,.
\end{equation}
Substituting Eqs~\eqref{k_x_prime} and \eqref{k_y_prime} into Eq.~\eqref{eq:second_order_derivative_operator_FT} and then performing inverse Fourier transform yields 
\begin{equation}
\label{eq:second_order_derivative_operator_1}
\mathscr{P}=(a^2\cos^2{\varphi}+b^2\sin^2{\varphi})\frac{\partial^2 }{\partial x^2}+(a^2-b^2)\sin{(2\varphi)}\frac{\partial^2 }{\partial x \partial y}+(a^2\sin^2{\varphi}+b^2\cos^2{\varphi})\frac{\partial^2 }{\partial y^2}\,.
\end{equation}
Since the stimulus $S$ is oriented along the $x'$ axis, we would need more weight on $\partial^2/\partial y'^2$ than on $\partial^2 /\partial x'^2$ to have a maximal response to the desired orientation; this implies that $b^2 \geq a^2$.

\subsection{Receptive field (RF) operator}
\begin{figure}[h!]
\centering
\includegraphics[width=.8\textwidth]{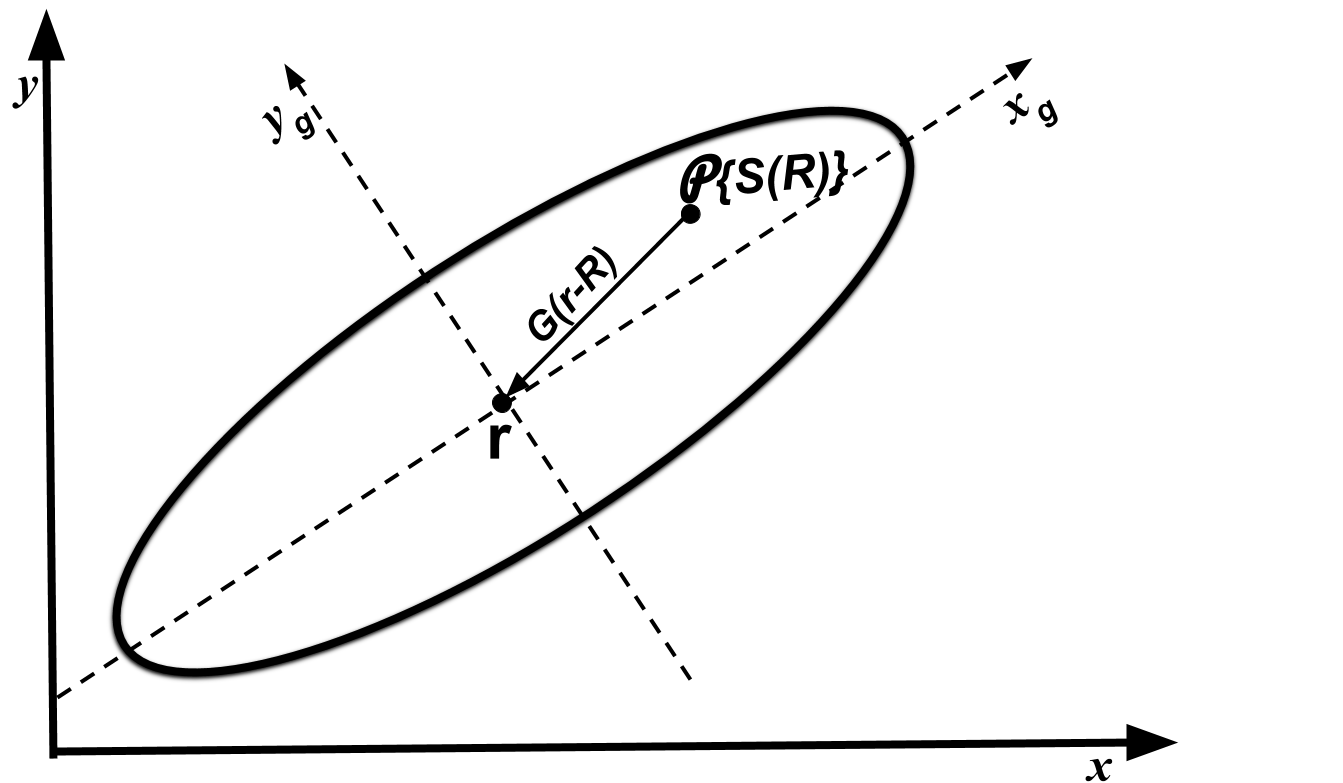}
\caption{Schematic of operators leading to the OP response at cells at $\mathbf{r}$ on the cortex. An oriented bar is mapped to locations $\mathbf{R}$ in the receptive field, which projects to {\bf r} via the anisotropic weight function $G(\mathbf{r}-\mathbf{R})$ indicated by the solid elliptic contour. The local anisotropic Laplacian operator $\mathscr{P}$ then acts at {\bf r}. The arrow shows how the neural response are project to measurement point $\mathbf{r}$ via the weight function. The $x_g$ and $y_g$ $\mathit{a}$ $\mathnormal{a}$ are the major and minor axis of the weight function, respectively.}
\label{fig:schematic_find_response_at_r}
\end{figure}

Previous studies \citep{ferster2000neural, finn2007emergence,hubel_ferrier_1977,marino2005invariant, toth1997integration,schummers2002synaptic,troyer1998contrast} have suggested that the orientation tuning of V1 neuron is altered by the excitatory (inhibitory) inputs from locally connected neighboring neurons (either from the same orientation column or from neighboring columns). Thus, we now introduce an RF operator to model the anisotropic RF that projects to V1 cells. This consists of a weight function $G(\mathbf{R}-\mathbf{r})$, which describes the strength of the neural projection from $\mathbf{R}$ to a cell at $\mathbf{r}$.  Figure~\ref{fig:schematic_find_response_at_r} shows the schematic of the RF operator on a piece of cortex. It shows the location $\mathbf{r}$ of the cells whose OP is being approximated and locations $\mathbf{R}$ whose activity projects to {\bf r}.

We approximate $G(\mathbf{r}-\mathbf{R})$ as an anisotropic Gaussian function whose long axis is oriented at the local OP $\varphi$ at $\mathbf{R}$ \citep{jones_gabor_RF}. If $\mathbf{R}=(x_{R},y_{R})$ and $\mathbf{r}=(x,y)$, we have
\begin{equation}
G(\mathbf{r}-\mathbf{R})= \frac{1}{2\pi \sigma_{x} \sigma_{y}}\exp \left [-\frac{1}{2}\left(\frac{x_{g}^{2}}{\sigma_{x}^{2}}+\frac{y_{g}^{2}}{\sigma_{y}^{2}}\right)\right ],
\label{eq:weight_fn}
\end{equation}
where
\begin{equation}
\label{x_g_axis}
x_{g}=(x-x_{R})\cos\varphi+(y-y_{R})\sin\varphi,
\end{equation}
\begin{equation}
\label{y_g_axis}
y_{g}=-(x-x_{R})\sin\varphi+(y-y_{R})\cos\varphi.
\end{equation}

The appropriate width of $G(\mathbf{r}-\mathbf{R})$ along the $x_{g}$ axis is determined by two factors: (i) the approximate RF size near the fovea measured from experiments. Previous studies \citep{dow1981magnification,hubel1974uniformity,keliris2019estimating_human_suRF} yielded an RF size of $\approx 0.082^\circ$ at the eccentricity of $1^\circ$ in macaque Monkey. We then transform the RF size in visual degree to the corresponding cortical size in mm by adopting the magnification factor (mm/deg) from \cite{horton1991_RFrepresentation}, where the magnification factor $M$ in Monkey in approximated as 
\begin{equation}
 M=\frac{12}{E+0.75},
\end{equation}
where $E$ is the eccentricity in degrees. The corresponding cortical RF size is then $\sim \SI{0.56}{mm}$;
(ii) the width of the weight function should ensure an approximate $30^\circ$ fall off from the measuring neuron's maximum response when it is activated by its optimal orientation \citep{DEVALOIS_visual,Moore2012DevelopmentOO, moshe_tuning_curve,Ringach5639, Swindale1998}. In order to satisfy both factors, we choose $\sigma_x = \SI{0.16}{mm}$.

\subsection{Combined OP operator}
Here we combine the anisotropic Laplacian and RF operators from above to obtain an overall OP operator and adjust its parameters to match experimental OP tuning curves.

The input at location $\mathbf{R}$ is approximated by applying the AL operator to the stimulus $S$ (i.e.,  $\mathscr{P}\{S(\mathbf{R})\}$ in Fig.~\ref{fig:schematic_find_response_at_r}), so that the local orientation sensitivity is picked out, while the weight function $G(\mathbf{R}-\mathbf{r})$ determines how much response from locations $\mathbf{R}$ are projected to $\mathbf{r}$. Hence, the response at $\mathbf{r}$ can be approximated by convolving the weight function with the OP operator on the stimulus at different location $\mathbf{R}$. It can be written as
\begin{equation}
I(\mathbf{r})=\int G(\mathbf{r}-\mathbf{R})\mathscr{P}\left\lbrace S(\mathbf{R})\right\rbrace d\mathbf{R}\,.
\end{equation}
In the Fourier domain, the convolution theorem yields
\begin{equation}
\label{eq:response_I_fourier}
I(\mathbf{k})=G(\mathbf{k})\mathscr{P}(\mathbf{k})S(\mathbf{k})\,.
\end{equation}
where the algebraic function $\mathscr{P}(\mathbf{k})$ is defined in Eq.~\eqref{eq:second_order_derivative_operator_FT}. We can thus  reverse the order of $\mathscr{P}(\mathbf{k})$ and $G(\mathbf{k})$ on the right hand side of Eq.~\eqref{eq:response_I_fourier} and inverse Fourier transform to obtain
\begin{equation}
\label{eq:nueral_response_at_r}
I(\mathbf{r})=\int  \mathscr{P}\left\lbrace G(\mathbf{r}-\mathbf{R})\right\rbrace S(\mathbf{R})d\mathbf{R}\,.
\end{equation}
Hence, the response at $\mathbf{r}$ becomes the convolution of a new combined OP operator $\mathscr{P}\left\lbrace G(\mathbf{r}-\mathbf{R})\right\rbrace$ with the stimulus itself.  This result agrees with previous studies \citep{graham1989visual,movshon1978spatial}, which indicated that V1 simple cell can be modeled as a linear filter and its responses are computed as the weighted integral of the Laplacian-transformed  stimulus, with the weights given by the RF pattern.

\begin{figure}[h!]
\centering
\includegraphics[width=0.5 \textwidth,keepaspectratio]{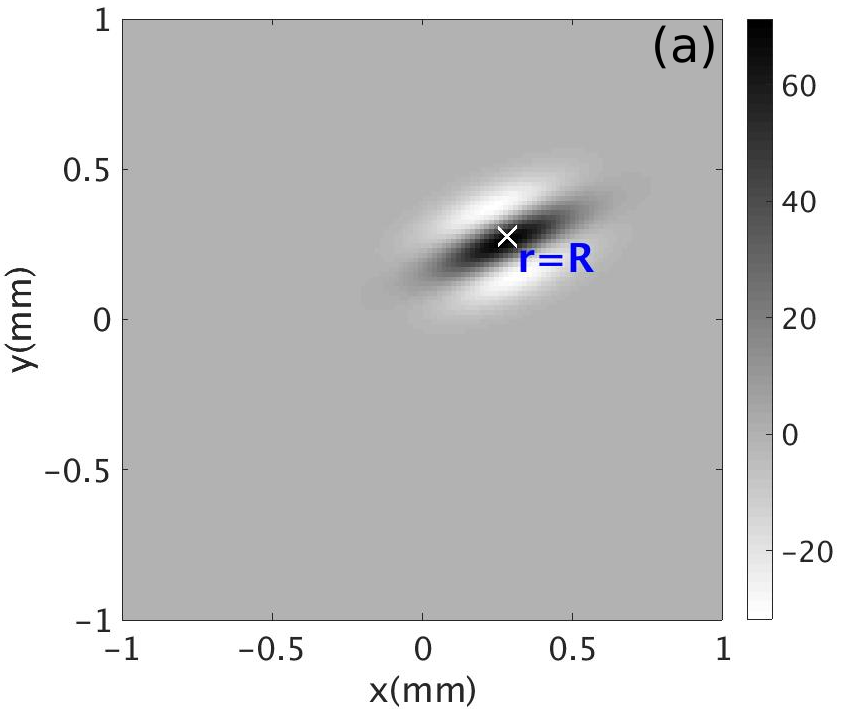}
\includegraphics[width=0.45 \textwidth,keepaspectratio]{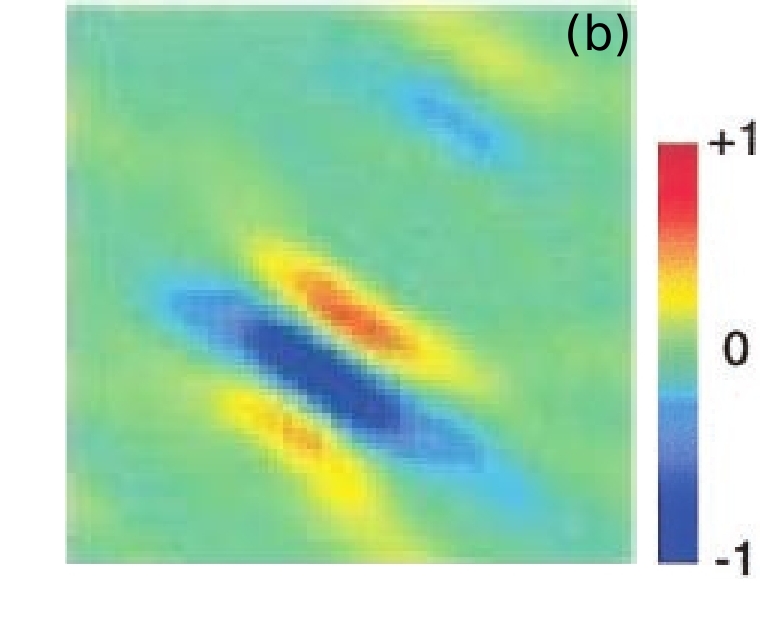}
\caption{(a) Receptive field operator $\mathscr{P}\left\lbrace G(\mathbf{R}-\mathbf{r})\right\rbrace$ with OP=$22.5^\circ$. The color bar indicates the amplitude of the operator. (b) RF of macaque monkey V1 simple cell from experiment \citep{ringach2002spatialRF}. The color bar indicates normalized impulse response strength of the neurons. }
\label{fig:operator_contour}
\end{figure}

Figure~\ref{fig:operator_contour}(a) shows a contour plot of the RF operator $\mathscr{P}\left\lbrace G(\mathbf{r}-\mathbf{R})\right\rbrace$ with a preferred orientation angle of $\varphi=22.5^\circ$. The operator has an elongated three-lobe pattern with its major axis along the direction $\varphi$, with an ON center lobe and two OFF side lobes. Figure~\ref{fig:operator_contour}(b) shows a measured RF of V1 simple cells of macaque monkey \citep{ringach2002spatialRF}, showing that our OP operator closely resembles the experimental one in spatial structure.  

\subsection{Angular selectivity of the OP operator}
\label{sec:operator_parra}
The full width at half maximum (FWHM) of the bell-shaped OP angle tuning curve plotted from the neuron response by convolving the RF operator and the stimulus can be used to parameterize the OP selectivity of the RF operator. The parameters are tunable by adjusting the ratio $\sigma_{x}^2/\sigma_{y}^2$ of the weight function  $G(\mathbf{r}-\mathbf{R})$ defined in Eq.~\eqref{eq:weight_fn}, and the ratio $b^2/a^2$ of the anisotropic Laplacian operator $\mathscr{P}$ defined in Eq.~\eqref{eq:second_order_derivative_operator_1}. Hence, we can find the optimal parameter values of the combined OP operator by adjusting its parameters so its tuning curve matches experiment.

\begin{figure}[h!]
\centering
\includegraphics[width=.9\textwidth]{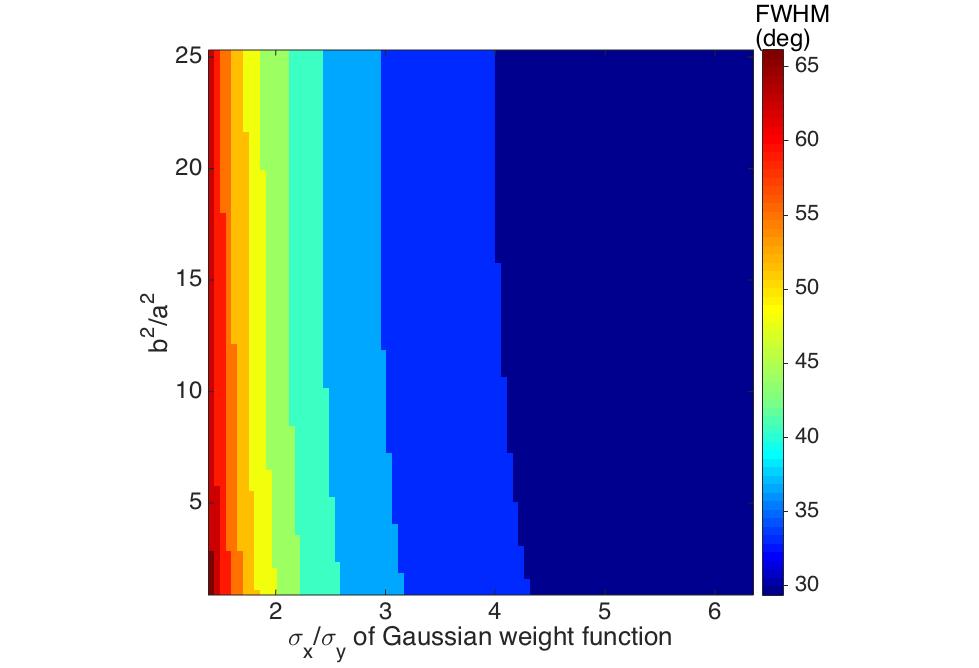}
\caption{Contour map of the FWHM of the OP tuning curve vs $b^2/a^2$, and $\sigma_{x}/\sigma_{y}$. The preferred orientation angle at measurement point is $135^\circ$. The value of $\sigma_{x}/\sigma_{y}$ is given by $x$ axis, and the ratio of $b^2$ and $a^2$ is given by $y$ axis. The color bar represents the FWHM width in degrees.}
\label{fig:angle_selctivity}
\end{figure}

We first vary the values of $a^2$ and $b^2$ while keeping their sum constant by writing
\begin{equation}
a^2=\sin^2{\psi}\,,
\end{equation}
\begin{equation}
b^2=\cos^2{\psi}\,,
\end{equation}
where $\psi$ ranges from 0 to $\pi/4$ to ensure $b^2\geq a^2$. We also vary the ratio $\sigma_{x}/\sigma_{y}$ from 1.5 to 6.5 to elongate the weight function along the $x_{g}$ axis defined in Eq.~\eqref{x_g_axis}. Figure~\ref{fig:angle_selctivity} shows the resulting contour map of the FWHM vs.~$b^2/a^2$ and $\sigma_{x}/\sigma_{y}$. The FWHM varies rapidly with $\sigma_{x}/\sigma_{y}$, with sharper tuning as $\sigma_{x}/\sigma_{y}$ increases. In contrast, the FWHM only sharpens slightly when $b^2/a^2$ increases.

\begin{figure}[h]
\centering
\includegraphics[width=0.47\textwidth]{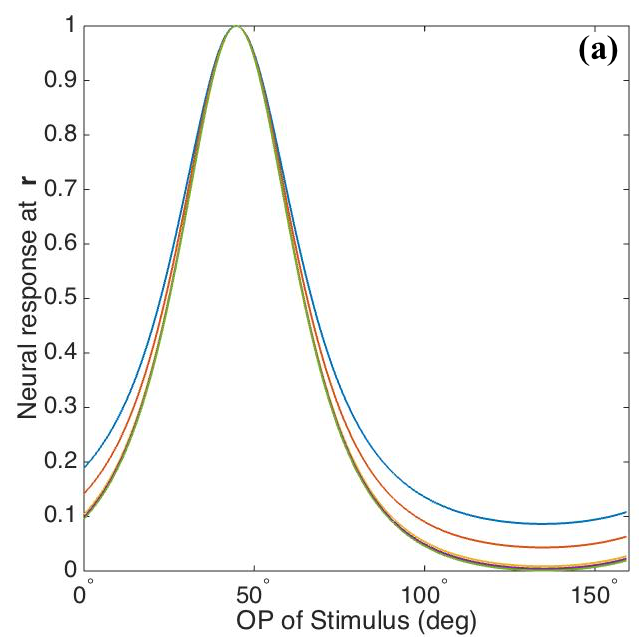}
\includegraphics[width=0.55\textwidth]{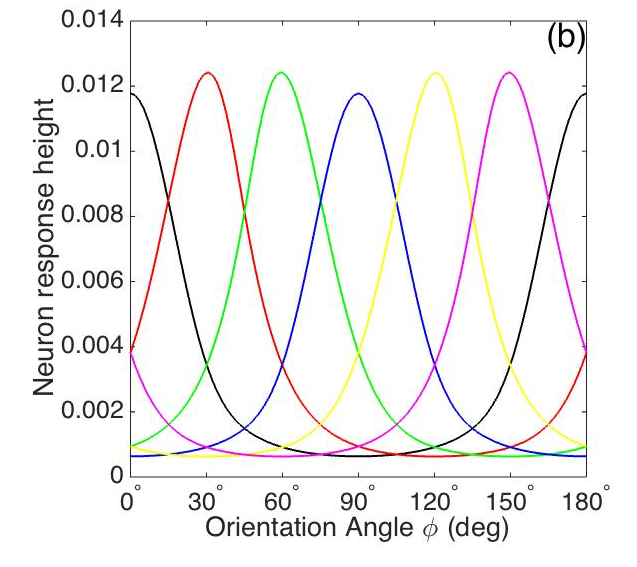}
\caption{(a) Normalized tuning curves with $b^2/a^2$ set to $1$ (blue), $2$ (red), $10$ (yellow), $20$ (purple), and $100$ (green); and fixing $\sigma_{x}/\sigma_{y}=2.5$. The preferred orientation angle is set to $45^\circ$. (b) OP tuning curves with orientation angles $0^\circ$ (black), $30^\circ$ (red), $60^\circ$ (green), $90^\circ$ (blue), $120^\circ$ (yellow), $150^\circ$ (purple), and $180^\circ$ (black), with $\sigma_{x}/\sigma_{y}=2.6$ and $b^2/a^2=1$. }
\label{fig:tuning_plot_norm}
\end{figure}

In order to illustrate the insensitivity of the FWHM to $b/a$ more  clearly, Fig.~\ref{fig:tuning_plot_norm}(a) shows the normalized tuning curves for fixed $\sigma_x/\sigma_y=2.5$, varying $b^2/a^2$ from 1 to 100. The FWHM decreases by only $\sim 2^\circ$ when $b^2/a^2$ changes from 1 to 5, and it does not decrease significantly further for larger $b^2/a^2$. The reason for this is that the RF operator envelope defined by  $G(\mathbf{r}-\mathbf{R})$ limits the effective lengths of its three lobes to the Gaussian envelope's characteristic width, so they do not change much when $b^2/a^2$ increases. This agrees with the predictions of previous studies, which argued that the OP tuning width of a V1 simple cell varies inversely with the size of its RF \citep{hubel1962receptive, lampl2001prediction}. It is also consistent with the Gaussian derivative model proposed by \cite{young2001gaussian} for modeling the spatiotemporal RF of V1 cells. Our results also match Hubel and Wiesel's feed-forward model, in which the overall V1 RF results from the net effect of aggregating isotropic LGN RFs via  anisotropic connections in V1 \citep{hubel1962receptive}. 

The above analysis implies that we can simplify the OP operator by setting $b^2=a^2$ because their ratio does not affect the OP tuning width significantly. Then the OP operator $\mathscr{P}\left\lbrace G(\mathbf{r}-\mathbf{R})\right\rbrace$ becomes the Laplacian of the weight function $\mathscr{L}\left\lbrace G(\mathbf{r}-\mathbf{R})\right\rbrace$ and the tuning width is controlled by the elongation of the weight function $G(\mathbf{r}-\mathbf{R})$.

Previous studies suggested that the FWHM of orientation tuning curve of most V1 neurons is $35^\circ$ to $40^\circ$ \citep{DEVALOIS_visual,moshe_tuning_curve,Ringach5639,swindale_review_1996}. This corresponds to $\sigma_{x}/\sigma_{y}$ ranging roughly from 2.3 to 3.2 in Fig.~\ref{fig:angle_selctivity}. Thus, to be consistent with experimental results, we choose $\sigma_{x}/\sigma_{y} = 2.6$ and $b^2=a^2$, which gives a FWHM of $37^\circ$ and the tuning curves shown in Fig.~\ref{fig:tuning_plot_norm}(b).

\subsection{Tuning curves vs.~Distance to Pinwheel Center}
\label{sec:tuning_curves_vs_distance}
The tuning curve of a cell (e.g., located at $\mathbf{r}_{0}$) describes its responses to different OPs. We compute the overall response of the cell for a particular OP [i.e., $I_{OP}(\mathbf{r}_{0})$] by taking a weighted average of the neural responses within a small circular region surrounding the cell and tightly coupled to it; this is achieved by integrating the responses of all the cells with a Gaussian weight function over the region:
\begin{equation}
   I_{OP}(\mathbf{r}_{0})= \int I(\mathbf{r})W(\mathbf{r}-\mathbf{r}_0)d\mathbf{r}\,,
\end{equation}
where
\begin{equation}
   W(\mathbf{r}-\mathbf{r}_0)=\frac{1}{2\pi \sigma_{r}^2}\exp \left [-\frac{(\mathbf{r}-\mathbf{r}_0)^{2}}{2\sigma_{r}^{2}}\right ]\,.
\end{equation}
The width of the weight function is set to 40 $\mu$m, to approximate the characteristic  width of an OP microcolumn and the experimental range of pinwheel-center effects on OP selectivity  \citep{maldonado1997orientation,nauhaus2008neuronal,obermayer_geometry_1993,ohki2006highlyorderpinwheel}. 
\begin{figure}[h!]
\centering
\includegraphics[width=0.7\textwidth]{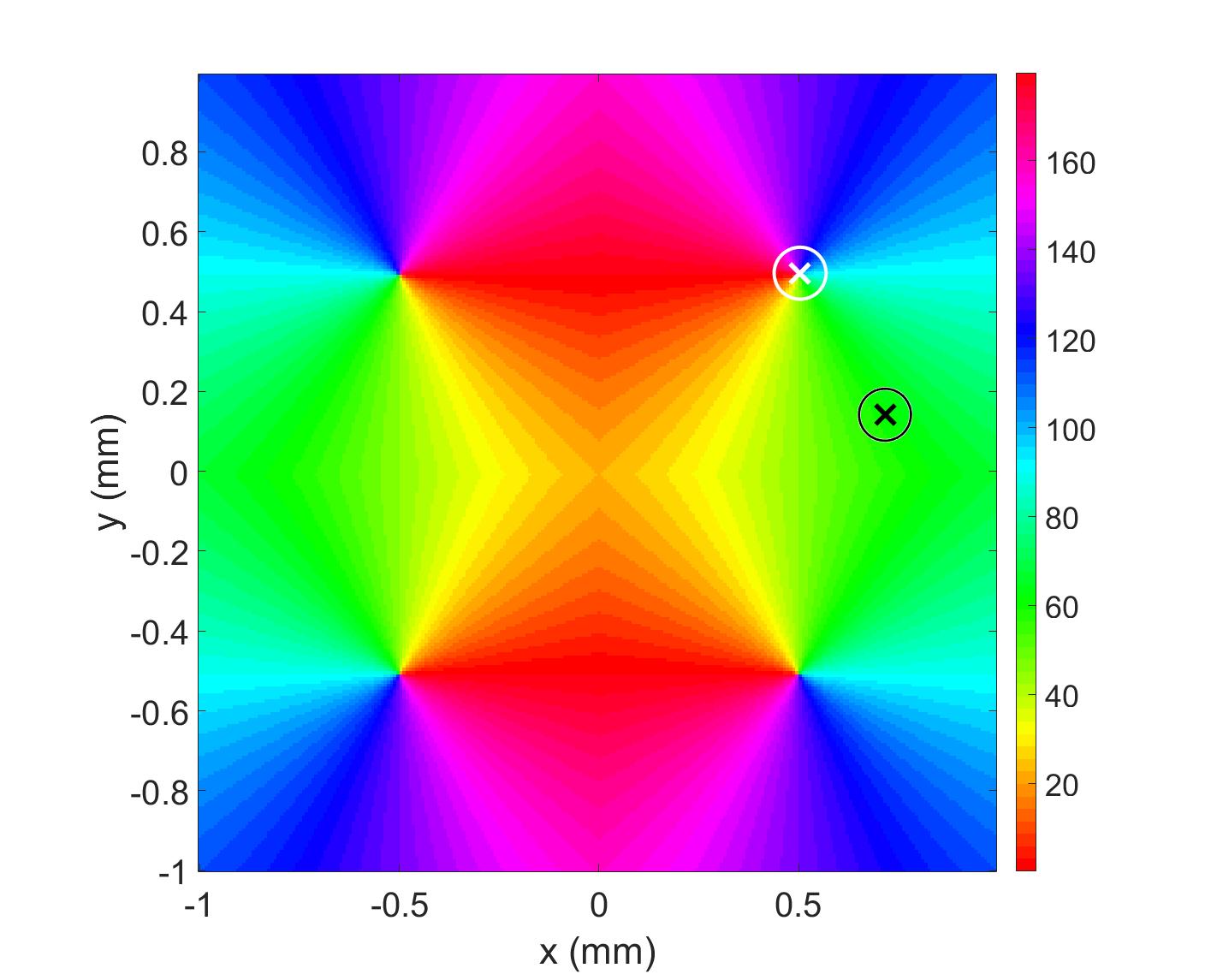}
\caption{Locations of cells in a hypercolumn for computing the tuning curves. Cells near a pinwheel center and in an iso-orientation domain are marked with crosses, and the circle around each  indicates the characteristic width of the integration region. The color bar shows the orientation angle in degrees.}
\label{fig:op_map_iso_pwc_location}
\end{figure}

\begin{figure}[h!]
\centering
\includegraphics[width=.47\textwidth]{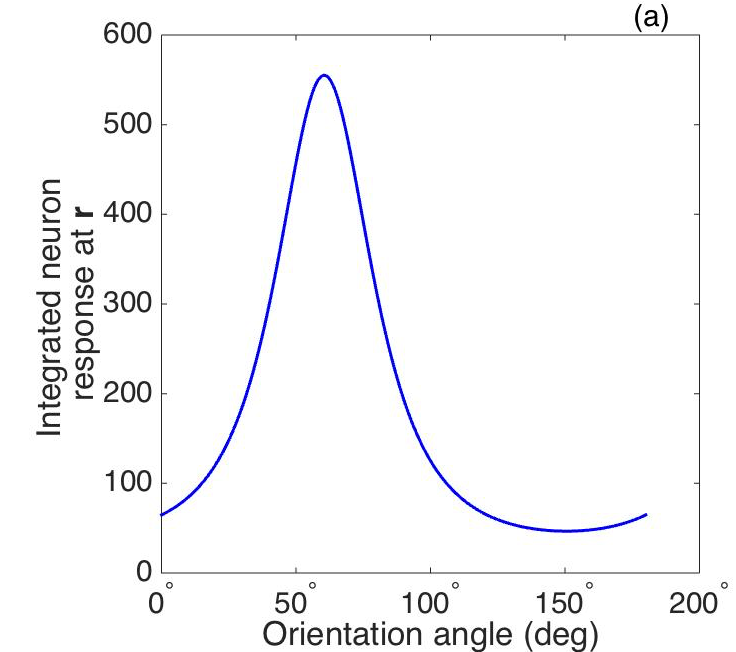}
\includegraphics[width=.45\textwidth]{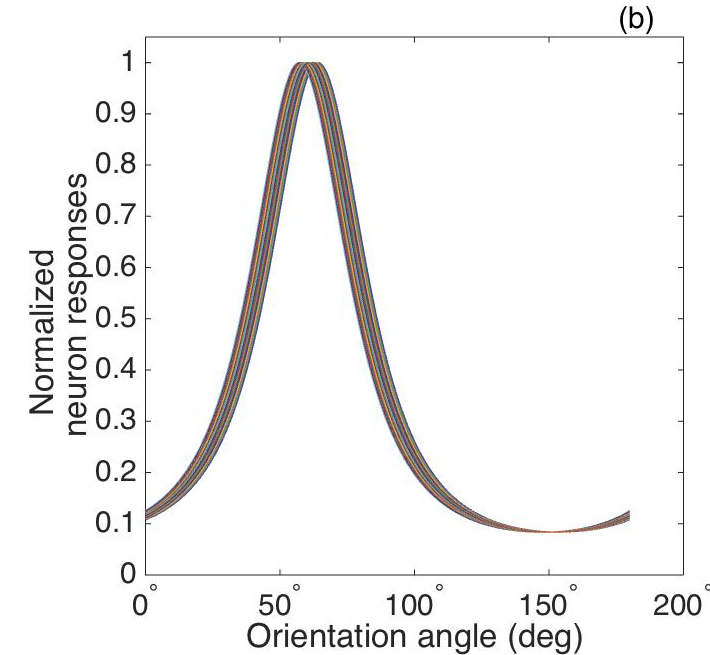}
\includegraphics[width=.47\textwidth]{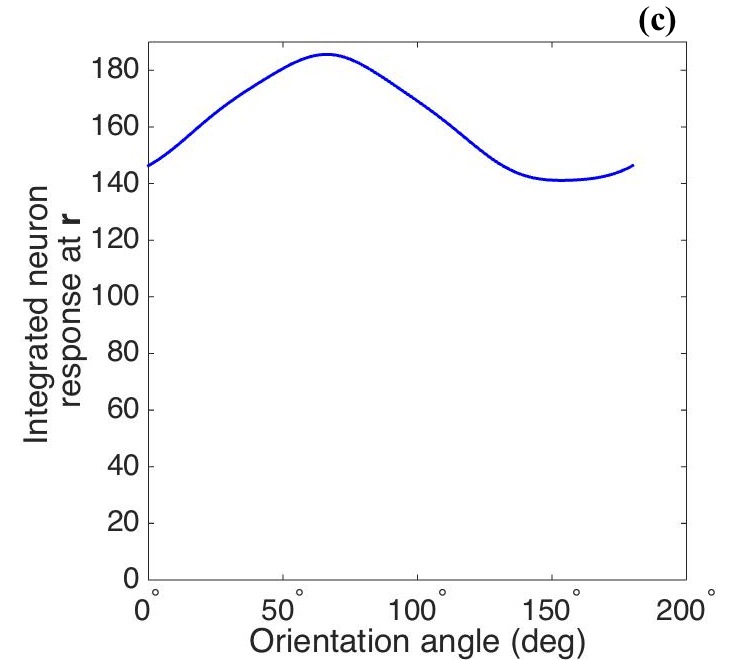}
\includegraphics[width=.47\textwidth]{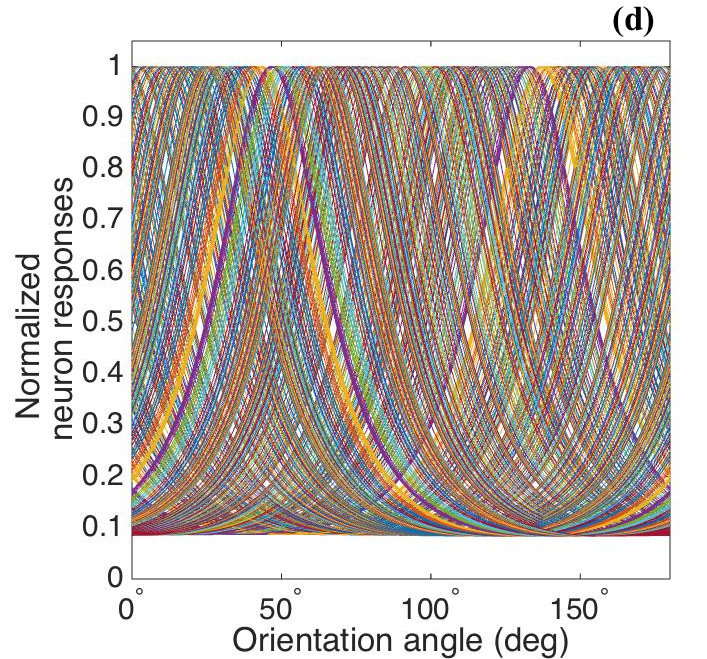}
\caption{(a) Tuning curve of averaged responses at measurement site located in iso-orientation domain (i.e. marked as black cross in Fig.~\ref{fig:op_map_iso_pwc_location}). (b) Tuning curves of all the cells surrounding the measurement site within the circular region in iso-orientation domain. (c) Tuning curve of averaged responses at measurement site located near pinwheel center (i.e. marked as white cross in Fig.~\ref{fig:op_map_iso_pwc_location}). (d) Tuning curves of all the cells surrounding the measurement site within the circular region near pinwheel center.}
\label{fig:tuning_curves_iso}
\end{figure}

We consider two cases of tuning curves for cells with different locations in the hypercolumn, one near a pinwheel center and another one in an iso-orientation domain. These locations are marked with crosses in Fig.~\ref{fig:op_map_iso_pwc_location}, and the circle around each cross indicates the characteristic width of the weight function in Eq.~(18). Figure~\ref{fig:tuning_curves_iso}(a) shows the resulting tuning curve of the overall responses of the cell located in an iso-orientation domain (i.e., the location marked with black cross in Fig.~\ref{fig:op_map_iso_pwc_location}) with preferred orientation $\approx 60^\circ$. It is sharply peaked at the preferred angle with FWHM $\approx 41^\circ$. In Fig.~\ref{fig:tuning_curves_iso}(b), we plot the tuning curves for an array of cells that are around the measurement site within the circular region. Since all the cells are located in the  iso-orientation domain, they have very similar orientation preferences and tuning curves. The overall response of the cell near the pinwheel center is plotted in Fig.~\ref{fig:tuning_curves_iso}(c), it is much broader than the tuning curve shown in Fig.~\ref{fig:tuning_curves_iso}(a) due to the fact that we average the responses from neurons with a wide range of OPs, as shown in Fig.~\ref{fig:tuning_curves_iso}(d). Our predictions agree with the experimental results \citep{maldonado1997orientation,ohki2006highlyorderpinwheel}, who found that individual neurons near the pinwheel center are just as orientation selective as the ones in the iso-orientation domain, and the overall broadly tuned response is the averaged response of nearby cells with a wide range of OPs.

\begin{figure}[h!]
\centering
\includegraphics[width=.5\textwidth]{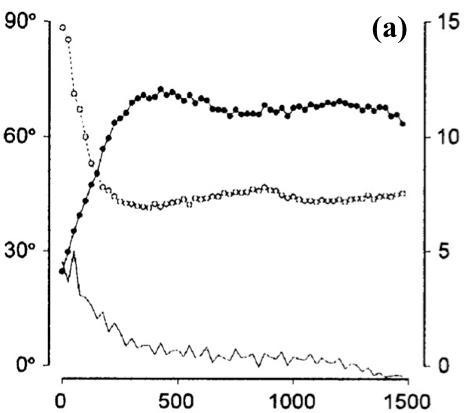}
\includegraphics[width=.55\textwidth]{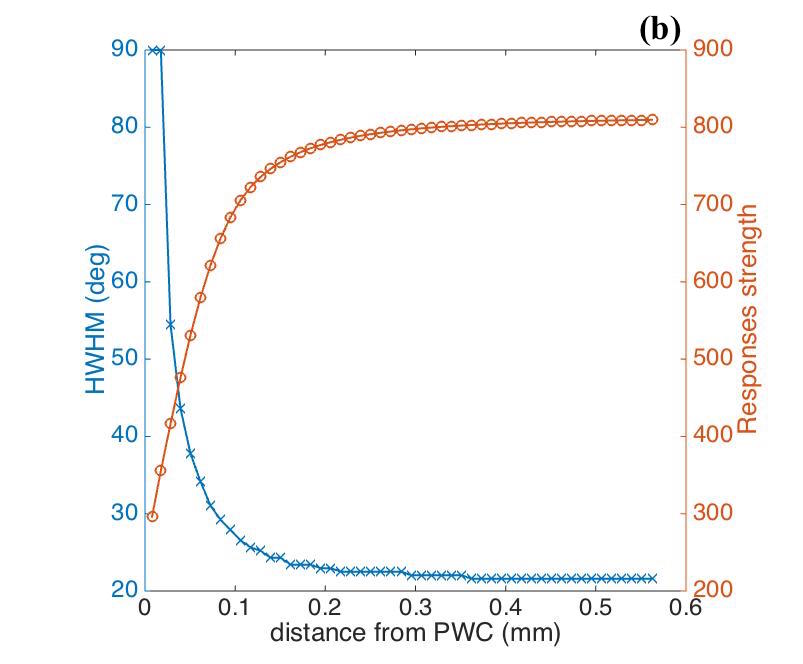}
\includegraphics[width=.55\textwidth]{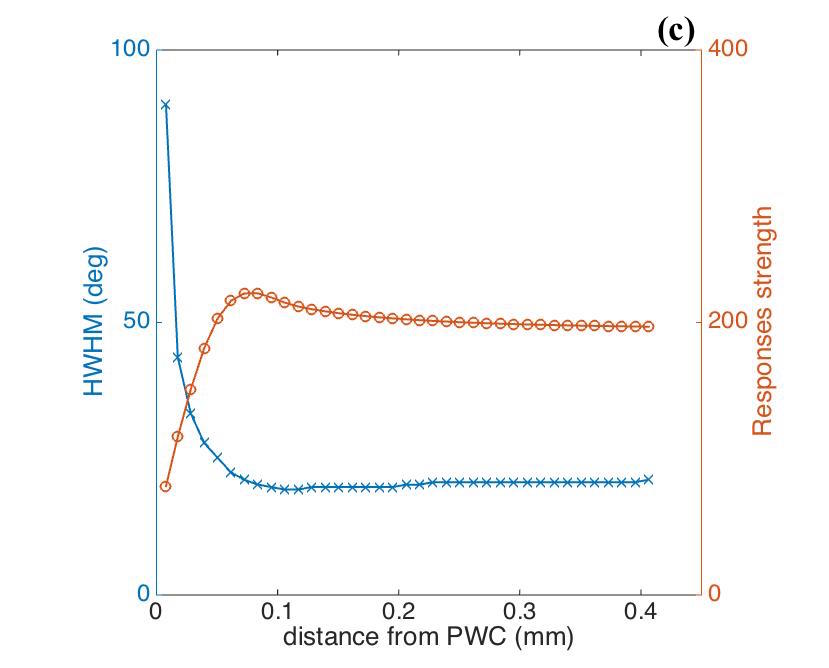}
\caption{(a) Experimental HWHM and response strength vs.~distance in $\mu$m from pinwheel center from \cite{swindale2003spatialpattern}, averaged over 13 pinwheels. The filled circles shows the response strength in arbitrary units, while the open circles show the HWHM in degrees and the bottom-most curve shows the baseline activity. (b) Predicted HWHM vs.~distance (blue) and Responses strength vs.~distance (orange) from pinwheel center, by using Gaussian function as weight function for averaging the responses. (c) Predicted HWHM vs.~distance (blue) and Responses strength vs.~distance (orange) from pinwheel center, by using Mexican hat function as weight function for averaging the responses.}
\label{fig:experiment_FWHM_curves}
\end{figure}

We have also investigated how the tuning width and response strength vary with distance from the pinwheel center, and compare them with experiment in Fig.~\ref{fig:experiment_FWHM_curves}. We calculate half width at half maximum (HWHM) here, in order to be consistent with the experimental plots. As expected, our predicted HWHM decreases when moving away from the pinwheel center to an iso-orientation domain, while the response strength increases with the distance (Fig.~\ref{fig:experiment_FWHM_curves}(b)). Both results match the experimental findings shown in Fig.~\ref{fig:experiment_FWHM_curves}(a), except the experimental HWHM in iso-orientation domain is wider than ours; However, our plots do not reproduce the overshoot and dip in the responses strength and HWHM curves, respectively, shown in Fig.~\ref{fig:experiment_FWHM_curves}(a). In order to achieve a better fit to the experimental data, we thus try the Mexican hat function,
\begin{equation}
   W_{mex}(\mathbf{r}-\mathbf{r}_0)=\left[1-\frac{(\mathbf{r}-\mathbf{r}_0)^{2}}{2\sigma_{r}^{2}}\right]\exp \left [-\frac{(\mathbf{r}-\mathbf{r}_0)^{2}}{2\sigma_{r}^{2}}\right ]\,.
\end{equation} 
as the weight function to average the responses, and the resulting plots are shown in Fig.~\ref{fig:experiment_FWHM_curves}(c). Overshoot and dip features are visible in this case, implying a better match. Thus, it is potentially possible to deduce the shape of the weight function from experimental results such as these, but detailed exploration is beyond the scope of the present paper.

\section{Fourier decomposition of the OP-OD Map}
\label{sec:fourier_analysis}
We seek a representation of the OP map in the Fourier domain so that we can apply it to compactly represent experimental data and to study the spatiotemporal neural activity patterns in periodic V1 structures using NFT, which requires such Fourier coefficients as input. Thus, in this section, we decompose the OP-OD map of the hypercolumn that is defined in Sec.~\ref{sec:unitcell} in the Fourier domain, derive the Fourier coefficients that represents the spatial frequency components of the OP-OD map structure and discuss their properties. We also determine the least number of Fourier coefficients we need in NFT analysis while maintaining the essential features of the OP-OD map. This is achieved by reconstructing the OP-OD map with a subset of the coefficients using the inverse Fourier transform.

We decompose the OP map in the hypercolumn by first applying a spatial operator $\mathscr{O}$ to the map, with $\mathscr{O}$ defined as
\begin{equation}
\label{eq:operator_o}
    \mathscr{O}=\exp{[i2\varphi(x,y)]}.
\end{equation}
This operator preserves the structure and periodicity of the OP-OD map and allows us to avoid the spurious discontinuities between $0^\circ$ and $180^\circ$ orientations, which actually correspond to the same stimulus orientation. This is important because representation of such discontinuities would require use of high spatial frequencies, and thus many Fourier coefficients. 
We then perform a 2D Fourier transform on the resulting map, which yields a sparse set of Fourier coefficients.

\begin{figure}[h!]
\centering
\includegraphics[width=0.6\textwidth]{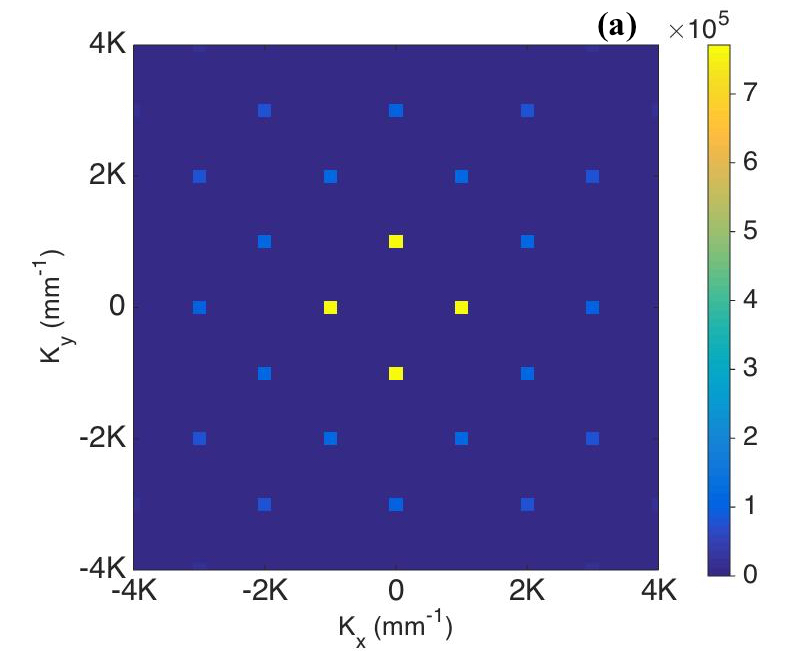}
\includegraphics[width=0.6 \textwidth]{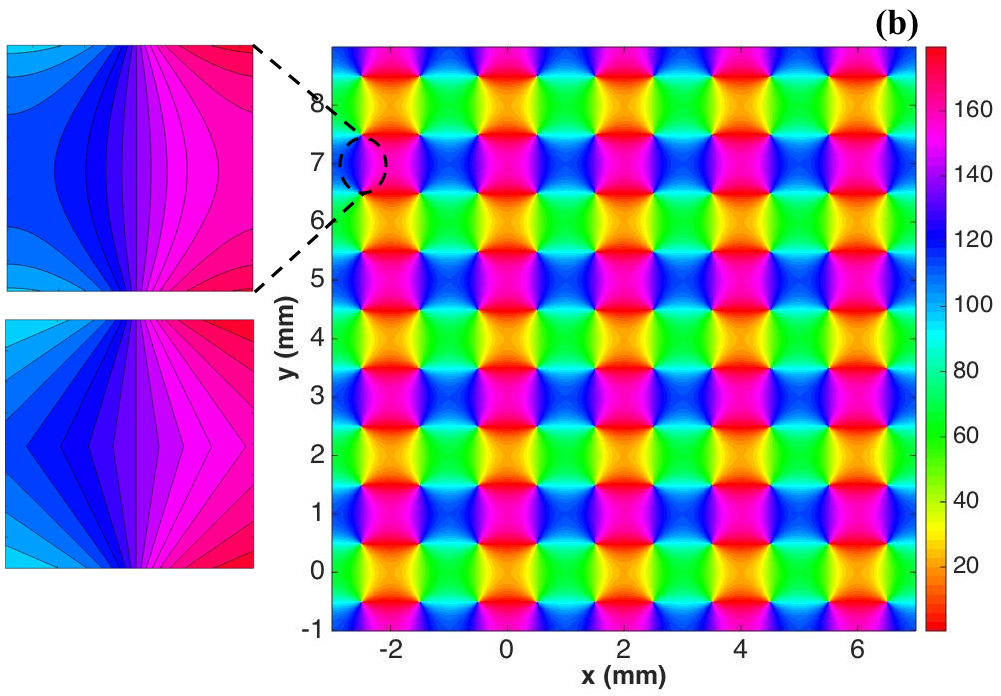}
\includegraphics[width=0.8 \textwidth]{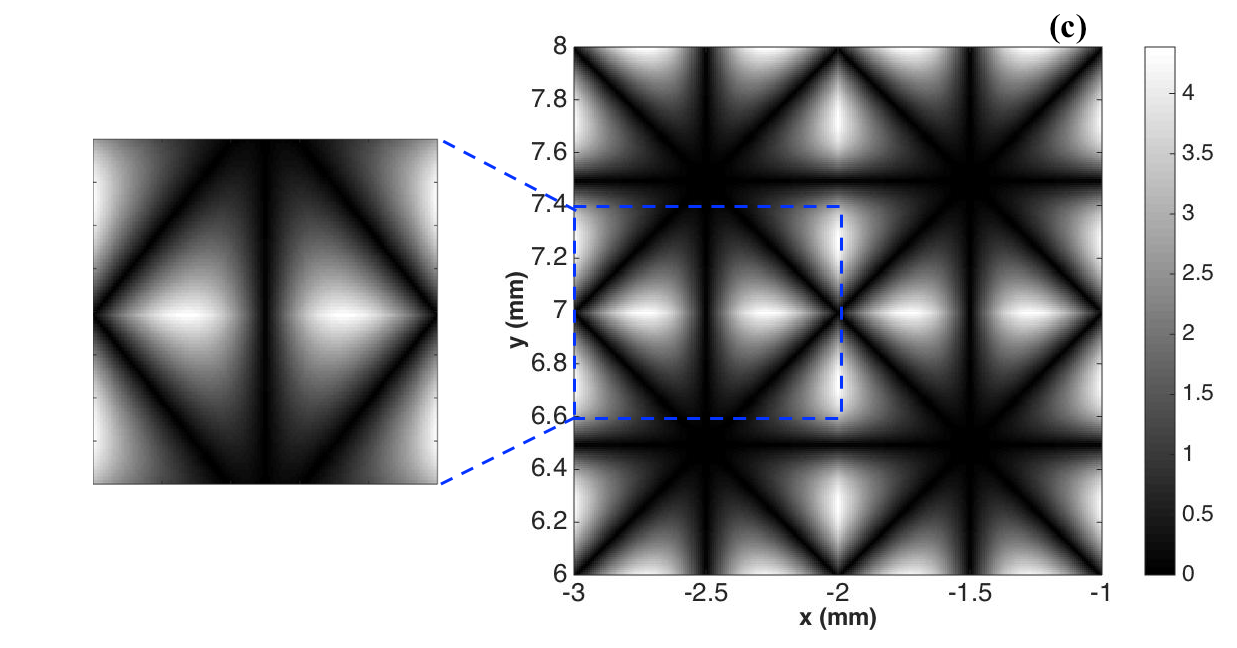}
\caption{(a) Magnitude of Fourier coefficients of the OP-OD map after applying the operator $\mathscr{O}$ to a lattice of 25 hypercolumns. Each square on the figure represents one spatial mode $\mathbf{K}$, and the color bar indicates the magnitude of it. (b) Reconstructed OP map of a lattice of 25 hypercolumns. The two squares on the top-left are the zoomed-in patches of the reconstructed (top) and the original(bottom) lattice. These are extracted from the same location that is marked by dashed-line oval. The color bar indicates the OP in degrees. (c) Absolute differences between the original hypercolumn OP in Fig.~\ref{fig:op_map}(c) and the reconstructed one. The square on the left shows a zoomed-in patch that are extracted from the location marked by blue dashed-line. The color bar indicates the difference in degrees.}
\label{fig:mag_plot_25_unitcell}
\end{figure}

Figure~\ref{fig:mag_plot_25_unitcell}(a) shows the magnitude of the Fourier coefficients of a lattice of $5\times5$ hypercolumns [i.e., Fig.~\ref{fig:op_map}(d)]. We note that: (i) the coefficients have 4-fold symmetry, and the 4 lowest $\mathbf{K}$ modes are dominant; and (ii) the lowest $\mathbf{K}$ modes are located at $(\pm \pi/a,0)$ and $(0,\pm \pi/a)$, where $2a$ is the width of hypercolumn. 

One of our main aims in finding these Fourier coefficients of the OP-OD map is to incorporate the OP map structure into the patchy propagator theory introduced to treat periodic V1 structure in previous studies \citep{robinson_patchy_2006,robinson_visual_2007, Lxc2020gamma_correlation}. We want to use as few coefficients as possible to simplify computation, while preserving the essential OP-OD structure. In order to test how well a small subset of Fourier coefficients can approximate the OP-OD map, we reconstruct the lattice of hypercolumns from these coefficients and compare it to the original one.  We first perform the inverse Fourier transform on a small subset of the coefficients. The resulting complex data represents the values that have been transformed from the OP angles after applying the operator $\mathscr{O}$ defined in Eq.~\eqref{eq:operator_o}. We then transform the complex data back to OP angles via
\begin{equation}
   e^{i\varphi(x,y)}= \cos[{\varphi(x,y)}]+i\sin[{\varphi(x,y)}],
\end{equation}
whence
\begin{equation}
   \varphi(x,y)= \tan^{-1} \left[\frac{\sin{\varphi(x,y)}}{\cos{\varphi(x,y)}} \right].
\end{equation}

We find that the four lowest $\mathbf{K}$ modes (the yellow squares in Fig.~\ref{fig:mag_plot_25_unitcell})(a) suffice to reproduce the main features of the hypercolumn lattice, and the reconstructed OP map is shown in Fig.~\ref{fig:mag_plot_25_unitcell}(b), which is very similar to the original one [i.e., Fig.~\ref{fig:op_map}(d)] except some angular contours are smoothed out due to the absence of higher-$\mathbf{K}$ modes. This detail is shown in the top left frame of Fig.~\ref{fig:mag_plot_25_unitcell}(b), which is to be compared with the frame below it, which is from the same part of the original lattice.

Figure~\ref{fig:mag_plot_25_unitcell}(c) shows the absolute differences between the original hypercolumn and the reconstructed one. The square on the left is a zoomed-in patch that are marked by the dashed-line square, and it is extracted in the same location as we do for the zoomed-in patch in Fig.~\ref{fig:mag_plot_25_unitcell}(b). The largest difference is $\approx 4.5^\circ$, and are around the edges of each pinwheel. Nevertheless, the basic structure and periodicity of the hypercolumn are all preserved in the reconstructed lattice. Thus, we can conclude that the 4 lowest $\mathbf{K}$ modes are sufficient for incorporating the OP map structure into NFT computations.

Going beyond idealized hypercolumns, we also can model more irregular and biologically realistic OP maps by adding more spatial modes around the 4 lowest $\mathbf{K}$ modes. We have noticed that the real OP-OD map does not have straight OD columns running vertically as we defined in Sec.~\ref{sec:unitcell}; rather, these columns are bent and oblique. In image processing, a rotation of Fourier coefficients produces a rotation of the image in spatial domain by the same angle \citep{ballard1982computervision}. Hence, we can alter the OP-OD map by modifying the modes in Fourier domain. 

\begin{figure}[h!]
\centering
\includegraphics[width=0.45\textwidth]{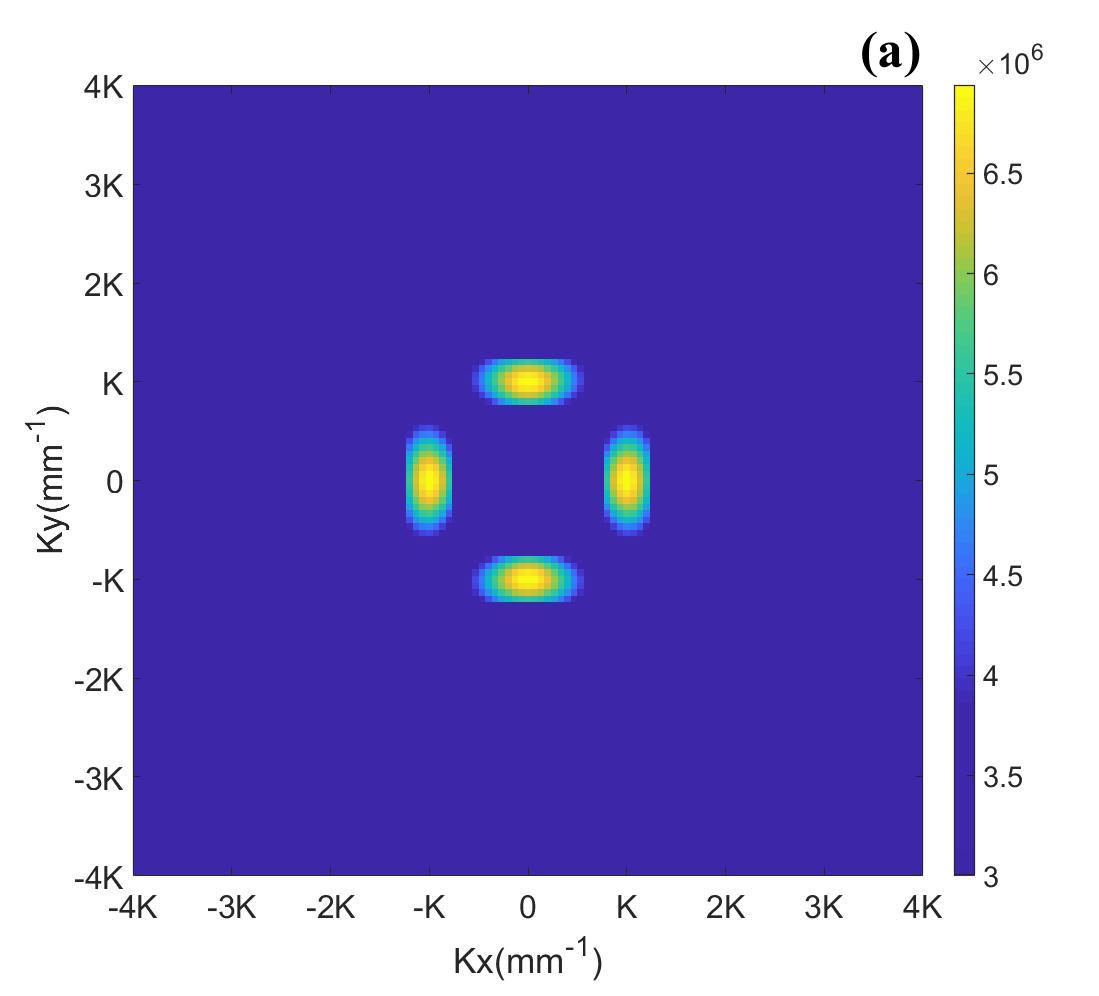}
\includegraphics[width=0.45\textwidth]{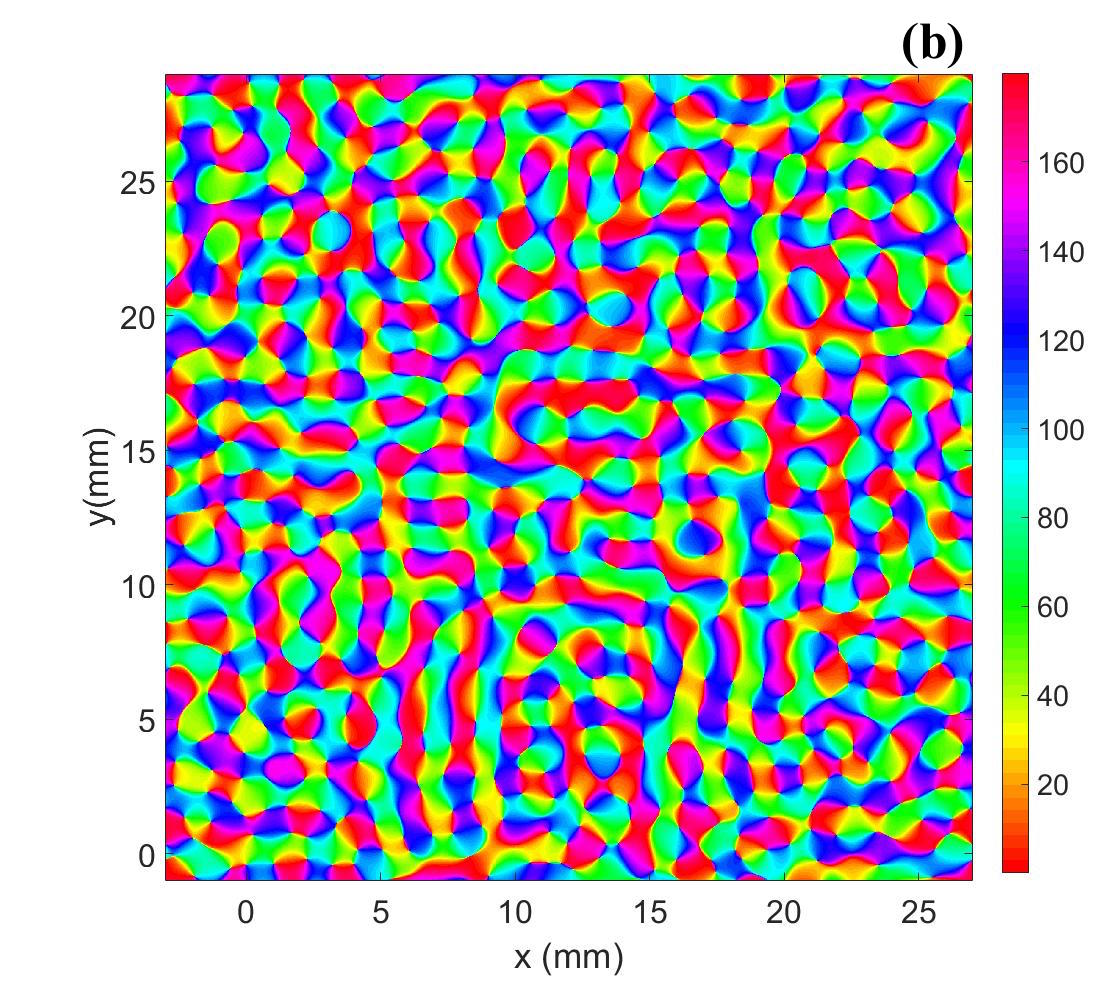}
\caption{Reconstructed OP-OD map with extra $\mathbf{K}$ in Gaussian envelope. (a) The magnitude plot of $\mathbf{K}$ modes. The color bar indicates the magnitude of it. (b) Reconstructed OP-OD map using set of $\mathbf{K}$ modes shown in (a). The color bar indicates the OP angle in degree. }
\label{fig:reconstruct_unitcell_Gaussian_K}
\end{figure}

Our approach for reconstructing the more realistic OP-OD map is to add the $\mathbf{K}$ modes around the 4 lowest $\mathbf{K}$  with its magnitude $M_{k}$ defined by Gaussian envelopes in $K_{x}$, $K_{y}$, and azimuthal angle $\Theta$, defined as,
\begin{equation}
   M_{k}= \exp\left [-(\mathbf{K}-\mathbf{K_{0}})^{2}/2\Delta_K^{2}\right ]\exp\left [-(\Theta-\Theta_{0})^{2}/2\Delta_\Theta^{2}\right ]\,,
\end{equation}
where $\mathbf{K_{0}}$ and $\Theta_{0}$ are the location and azimuth angle of the original lowest $\mathbf{K}$ modes, $\Delta_K$ and $\Delta_\Theta$ are the variance of the Gaussian envelope and we set these to $\frac{1}{2}K$ mm$^{-1}$ ($K=\pi/a$, where $2a$ is the width of the hypercolumn) and $20^\circ$, respectively, to match experimental observations. The resulting magnitude plot of $\mathbf{K}$ modes is shown in Fig. \ref{fig:reconstruct_unitcell_Gaussian_K}(a) and reconstructed OP-OD map using this set of $\mathbf{K}$ is shown in Fig.~\ref{fig:reconstruct_unitcell_Gaussian_K}(b).

 Fig.~\ref{fig:reconstruct_unitcell_Gaussian_K}(b) resembles the biologically realistic OP-OD maps obtained in experiments \citep{blasdel_orientation_1992,bonhoeffer_layout_1993,obermayer_geometry_1993}, and it reproduces the general observations of the OP-OD maps we mentioned in Sec.~\ref{sec: unitcell_arrangemt}, including: (i) it has both positive and negative pinwheels, and the neighboring pinwheels have opposite signs; (ii) linear zones connect two pinwheel centers.

\section{Application to OP maps from a neural network model}
\label{sec:applications}
In order to analyze the general properties of more realistic OP maps, We perform the same Fourier analysis as on the idealized hypercolumns in previous sections for the OP-OD map generated from a computational neural network model.

\begin{figure}[h!]
\centering
\includegraphics[width=0.45\textwidth]{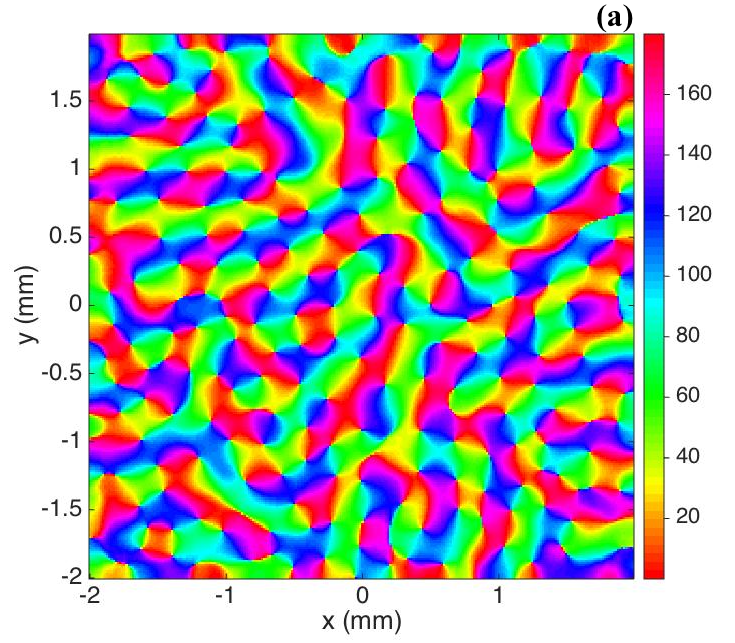}
\includegraphics[width=0.45\textwidth]{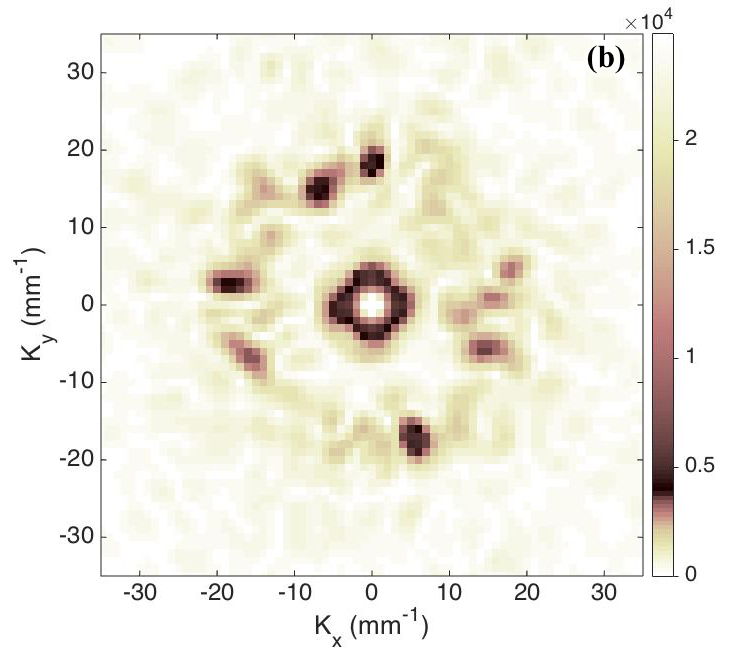}
\caption{(a) OP map generated from GCAL model \citep{bednar2009topographica}, and the color bar indicates the OP angle in degrees. (b) Magnitude plot of the Fourier coefficients obtained from GCAL OP map. Each pixel-like square represents one spatial mode $\mathbf{K}$, and the color bar indicates its magnitude.}
\label{fig:gal_op_map}
\end{figure}

The model of V1 we use here is the Gain Control, Adaptation, Laterally Connected (GCAL) model \citep{Stevens_GCAL}, which treats the retina, LGN, and V1 as 2-dimensional sheets, with neurons in each sheet connected topographically. Neurons not only connect to a small group of neurons of the lower level sheet, but also laterally connect to the neurons within the same sheet. A Hebbian learning rule is adopted in the model for updating the connection weights between neurons \citep{Stevens_GCAL}. Figure~\ref{fig:gal_op_map}(a) shows an example output from the GCAL model simulation \citep{bednar2009topographica}, and we use this map for further analysis in the Fourier domain. 

We apply the operator $\mathscr{O}$ to the GCAL OP map and then do a Fourier transform. One thing worth mentioning here is that the OP  has discontinuities at the edges of the map if we repeat the whole map in $x$ or $y$ direction, which is implicit in the 2-D discrete Fourier transform. This introduces edge effects into the Fourier transform, which we minimize by doubling the linear dimensions of the array and zero padding the added region. Figure~\ref{fig:gal_op_map}(b) shows the Fourier coefficients after performing Discrete Fourier transform on the zero padded map. The dominant $\mathbf{K}$ terms are shown in black. Except the ring-shaped enhancement near the center, which arises from the zero padding of the map and the linear size of the overall simulation area. The other dominant $\mathbf{K}$ terms correspond to the periodicity of the hypercolumns and have square symmetry similar to the pattern in Fig.~\ref{fig:mag_plot_25_unitcell}(a), but with $\sim \pm 15^\circ$ spread of $\mathbf{K}$. The square symmetry is most likely at least partly due to a combination of: (i) the artifacts introduced by the approximately square OP-OD unit cell, and (ii) a structural bias introduced by the square grid on which the GCAL model is simulated.

\section{Summary and Conclusion}
\label{sec:summary}
In this paper, we present an analytic description of the OP-OD map in V1. It includes modeling both the local neuron sensitivity to the stimulus orientation and the weighted projection from nearby neurons. The results and analysis includes:

(i) we approximate the periodic OP-OD map in a square grid of hypercolumn with parallel left and right OD columns with equal width and OP pinwheels with alternating signs. The approximation is idealized but are good enough for preserving the basic structure of the OP-OD map.

(ii) We propose an AL operator to detect the orientation of the stimulus for local neuron. It is a weighted sum of second order partial derivatives.

(iii) The OP operator reproduces the spatial arrangement of the receptive field of V1 simple cells. It is derived by finding the neuron responses by combining the AL operator with a weighted sum of the projections from neighbouring neurons. We optimize the parameters of the operator by controlling the width and angle selectivity of the response tuning curve, and we find that the orientation tuning is only affected significantly by the aspect ratio of the weight function, not the weights of the second order partial derivatives for detecting input orientation in local neuron. The orientation tuning sharpens when we elongate the OP operator along the orientation axis, in accord with experiment  \citep{hubel1962receptive, lampl2001prediction}. 

(iv) We account for the lower OP sensitivity and lower response strength near pinwheel centers by averaging OP over the characteristic microcolumn scale of 40 $\mu$m --- near centers many different OPs are averaged together, broadening the tuning curve. A Mexican hat function gives a better match to experiment, raising the possibilty of using such experimental results to infer the microscopic connectivity profile.

(v) Fourier domain analysis of the OP-OD maps were performed to generate a set of Fourier coefficients for compact representation and especially for use in NFT. Only the lowest $\mathbf{K}$ modes are needed to described the spatial structure of the hypercolumn. This simplifies the computational work when we integrate the OP map into the patchy propagator and investigate the neuron activities in V1 using NFT. Moreover, if we keep the lowest $\mathbf{K}$ modes as the basic modes and add extra modes around it using Gaussian envelopes, we could reconstruct a more realistic OP-OD map that is similar to the ones obtained from experiments or computer simulations. 

(v) We also perform Fourier analysis on the irregular OP-OD map generated by GCAL model. The dominant $\mathbf{K}$ modes have approximately square symmetry as an artifact of the square grid on which the simulations are done. 

Overall, we have succeeded in obtaining compact representations of an idealized joint OP-OD map, in both coordinate and Fourier spaces. Notably, the elongated generalized Gaussian operator dominates in determining OP and mutual consistency OP and OD maps strongly constrains the possible combined maps in hypercolumns, with four pinwheels, not one, required periodic unit of the hypercolumn lattice.

\section*{Acknowledgements}
This work was supported by the Australian Research 
Council under Laureate Fellowship grant FL1401000025, 
 Center of Excellence grant CE140100007, 
and Discovery Project grant DP170101778.

\bibliography{shorttitle,mybib}

\end{document}